\documentclass[onecolumn,amsmath,amssymb,nofootinbib,12pt]{article}
 
 \ifx\pdfoutput\not\undefined
 \pdfoutput=1
\fi
\usepackage{jheppub}
\usepackage{ifpdf}
\usepackage{mdframed}
\usepackage{comment}
\usepackage{graphicx,subcaption}
\graphicspath{ {figures/} }
\usepackage{epsfig}
\usepackage{pdfpages}
\usepackage{bbm}
\usepackage{graphicx,epstopdf}
\usepackage[numbers,sort&compress]{natbib}
\usepackage[makeroom]{cancel}
\usepackage{hyperref}
\hypersetup{pdftex,colorlinks=true,allcolors=blue}
\usepackage{array}
\usepackage[export]{adjustbox}
\usepackage{amsthm}
\usepackage[normalem]{ulem}

\newtheorem{thm}{Theorem}

\theoremstyle{definition}
\newtheorem{defn}[thm]{Definition}

\newcommand{\hatit}[1]{\hat{\mathcal{#1}}}
\newcommand{\tp}{\tilde{\partial}}
\newcommand{\ew}[1]{e_{W}(#1)}
\newcommand{\cw}[1]{c_{W}(#1)}
\newcommand{\E}[1]{\mathcal{E}_{#1}}
\newcommand{\area}[1]{\operatorname{area}\left(#1 \right)}

\usepackage[utf8]{inputenc}

\usepackage{tocloft}
\usepackage{bbold}
\usepackage{tikz}
\usetikzlibrary{backgrounds}
\usetikzlibrary{patterns}
\usetikzlibrary{arrows.meta}
\usetikzlibrary{shapes,snakes}
\tikzstyle{bag} = [align=center]
\usetikzlibrary{decorations.pathmorphing}
\def\bea{\begin{eqnarray}}
\def\eea{\end{eqnarray}}

\usepackage{hyperref}
\usepackage{subcaption}

 \newcommand{\badat}{\begin{alignedat}}
 \newcommand{\eadat}{\end{alignedat}}
 
 \def\be{\begin{equation}}
\def\ee{\end{equation}}

\def\p{\partial}

\usepackage{xcolor}

\newcommand{\pink}[1]{\textcolor{\pink}{#1}}

\definecolor{dblue}{rgb}{0.2,0.50,0.80}

\def\C{\mathcal{C}}

\def\A{\mathcal{A}}
\def\D{\mathcal{D}}

\def\X{\mathcal{X}}

\def\Y{\mathcal{Y}}

\def\V{\hatit{V}}
\def\W{\hatit{W}}

\def\scri{\mathcal I}

\DeclareFontFamily{OT1}{pzc}{}
\DeclareFontShape{OT1}{pzc}{m}{it}{<-> s * [1.10] pzcmi7t}{}
\DeclareMathAlphabet{\mathpzc}{OT1}{pzc}{m}{it}

\definecolor{vert}{rgb}{0.1367 0.543 0.1367}

\title{
Generalized Entanglement Wedges 
and the Connected Wedge Theorem
 }

\author[a]{Athira Arayath}
\author[b]{and Sabrina Pasterski}

\affiliation[a]{Leinweber Institute for Theoretical Physics and Department of Physics,
University of California, Berkeley, California 94720, U.S.A.} 
\affiliation[b]{Perimeter Institute for Theoretical Physics, Waterloo, ON N2L 2Y5, Canada}

\emailAdd{athira@berkeley.edu}
\emailAdd{spasterski@perimeterinstitute.ca}

\date{\today}

\abstract{
We use the framework of generalized entanglement wedges to revisit the connected wedge theorem (CWT). This construction identifies an entanglement wedge associated for any bulk region and allows us to rephrase the CWT in terms of the entanglement entropies of bulk regions. We establish new upper and lower bounds on the mutual information of boundary decision regions in terms of the entropies of certain bulk regions associated with a scattering configuration. We then define new bulk decision regions for which we show that a non-empty scattering configuration implies a connected entanglement wedge. This generalization of the CWT extends to asymptotically flat spacetimes.
}

\begin{document}

\maketitle

\section{Introduction}\label{sec:intro}
A core principle in the emergence of spacetime in AdS/CFT is that quantum information quantities are encoded in bulk geometry. This is illustrated by the connected wedge theorem (CWT) which relates causal structure in AdS$_3$ to the entanglement structure of the CFT state on the conformal boundary \cite{May:2019yxi,May:2019odp}. It states that the existence of a local scattering region in the bulk implies $O(1/G_N )$ mutual information between certain boundary regions associated with the scattering configuration.

The main idea -- that entanglement on the boundary can mediate scattering in the bulk -- seems more general than the AdS$_3$ context in which it was proven. However, when we attempt to generalize this discussion to asymptotically flat spacetimes, we immediately run into a few puzzles. First, we don't have a good handle on the dual theory on the boundary. While there has been a significant amount of work building up the AFS/BMSFT dictionary for 3D asymptotically flat gravity~(see~~\cite{Riegler:2016hah,Oblak:2016eij,Bagchi:2025vri} for pedagogical treatments and references therein), the status of even the analog of the RT proposal is surprisingly nuanced~\cite{Bagchi:2014iea,Jiang:2017ecm,cmp}. In particular, the causal structure of null infinity is degenerate, with the boundary undergoing a Carrollian limit wherein the theory becomes ultralocal~\cite{Duval:2014uva, Bagchi:2016bcd, Ciambelli:2018wre}. 

Specifically, under the flat contraction of AdS~\cite{Compere:2019bua,Compere:2020lrt,Hijano:2019qmi} that takes the boundary cylinder to the conformal boundary of Minkowski space, only two small bands near $\tau=\pm\frac{\pi}{2}$ map to $\scri^\pm$, while the intermediate times map to spatial infinity which is singular in the conformal compactification itself. Thus, even barring understanding the dynamics of the theory and figuring out how to appropriately evaluate the entanglement entropies of the boundary regions, we will see that the data identifying the boundary regions becomes degenerate unless we introduce some cutoff screen.  Some progress generalizing the CWT to dS$_3$ using screens can be found in~\cite{Franken:2024wmh}. Here we will start with a rather different approach, and then apply it to a screen-based generalization of the CWT in AdS that will survive the flat limit.

Our goal here is to reformulate the connected wedge theorem in a manner amenable to a flat space limit. Recently, Bousso and Penington~\cite{Bousso:2022hlz,Bousso:2023sya} have proposed an extension of the entanglement wedge prescription which defines an entanglement wedge $e(a)$ for any gravitational {\it bulk} region $a$. This is a fascinating generalization of the notion of holography in that it lets us phrase things like the entanglement wedge of a region, or whether two regions are in a connected phase, etc. purely from the bulk perspective. Due to the aforementioned subtleties with the conformal boundary in flat space, one can thus be tempted to use this framework to understand a flat analog of the CWT.
Namely by recasting the statements in connected wedge theorem in terms of these purely bulk quantities, we can hope to identify an analog that extends nicely to asymptotically flat spacetimes.  

This paper is organized as follows. We start by reviewing some preliminaries on the connected wedge theorem and generalized entanglement wedges in section~\ref{sec:prelim}. In section~\ref{sec:scatreg} we establish new bounds on the mutual information of the boundary regions $I(\V_1:\V_2)$ in the usual $2\rightarrow 2$ connected wedge set up, in terms of entropies of bulk scattering regions. In section~\ref{sec:bulkcwt} we propose a way to generalize the CWT that moves the $\V_i$ into the bulk. 
We establish various features that persist from the boundary version before using this construction to examine the flat limit in section~\ref{sec:flat}.
We close with some discussion of interpretations and next steps in section~\ref{sec:disc}.
Additional proof steps relevant to establishing these results can be found in the appendix. 

Our main results are: bounding the mutual information in terms of scattering region entropies 
\begin{equation}
    S_{gen}(e_{max}(s_{pts}'')) \leq I(\hatit{V}_{1}; \hatit{V}_{2}) \leq S_{gen}(e_{min}(s_{ent})) 
\end{equation}
and generalizing the AdS connected wedge theorem to a version where the decision regions $\hatit{V}_i$ are moved into the bulk. Namely, we identify bulk decision regions $v_i$ for which
\be\label{sptsconn}
s_{pts}\neq\emptyset~\Rightarrow~ e_{min}(v_1\Cup v_2)~{\rm connected}.
\ee
See equations~\eqref{bound} and~\eqref{sptsconn} below. Two interesting observations arise from these constructions: First, we are tempted to provide a more physical interpretation of the lower bound on the mutual information, in relation to the information being communicated in the corresponding quantum task. Second, the manner in which we are defining entanglement wedges for regions on the screen is a bit different than in other places~\cite{Lewkowycz:2019xse,Franken:2024wmh} and seems like a useful application of the generalized entanglement wedge construction.

 We will elaborate on these points further in the discussion section. Before getting started, let us establish some notation and conventions. Readers familiar with the connected wedge theorem and generalized entanglement entropies can skip to section~\ref{sec:scatreg}.

\subsection{Summary of Notation}
 Let $M$ denote a globally hyperbolic Lorentzian manifold (the bulk), and let $\tilde{M}$ denote its conformal completion so that the conformal boundary is denoted by $\tp M :=
 \partial \tilde{M}$. 
\begin{itemize}
    \item Hatted capital letters are used to denote regions of the conformal boundary. We will use italicized letters (e.g. $\hatit{A}, \hatit{B}, \hatit{C},...$) to denote codimension 0 regions, and plain letters (e.g. $\hat{{A}}, \hat{B}, \hat{C},...$) to denote codimension 1 regions.
    
    \item Plain lowercase letters (e.g. $a,b,c,..$) are used to denote codimension 0 bulk spacetime regions.
    
    \item Italicized (unhatted) capital case letters (e.g. $\mathcal{A}, \mathcal{B}, \mathcal{C}, ...)$ are used to denote surfaces (codimension 1 or greater) in the bulk.
    
    \item The causal future/past $J^{\pm}$, future/past domain of dependence $D^{\pm}$, and horizons $H^{\pm}$ are defined as in \cite{Wald:1984rg}. A hat indicates their restrictions to the boundary e.g. $\hat{J}^{\pm}$ denotes the causal future/past restricted to the boundary.  
    
    \item $\text{cl}(a) = a \cup \partial a$ denotes the closure of a set $a$.  $\partial a$ denotes the boundary of $a$ as a subset of $M$ whereas $\delta a$ denotes the boundary of $a$ as a subset of the conformal completion $\tilde{M}$. The part contained in the conformal boundary is denoted $\tp a := \delta a \cap \partial \tilde{M}$. 
\end{itemize}
When $M$ is an asymptotically AdS spacetime, we let $\hat{A}$ be a  codimension-1 boundary spatial region, and let $\hatit{A} = \hat{D}(\hat{A})$ be its domain of dependence. Whenever it is clear, we will we refer to $\hatit{A}$ and $\hat{A}$ interchangeably. 
\begin{itemize}
    \item $\ew{\hatit{A}}$ denotes the (original) entanglement wedge of a boundary region $\hatit{A}$ and $\E{\hatit{A}}$ denotes the corresponding HRRT surface.  
    \item $\cw{\hatit{A}}:= J^{+}(\hatit{A}) \cap J^{-}(\hatit{A})$ denotes the causal wedge of boundary region $\hatit{A}$. 
\end{itemize}
We will stick to 3D throughout and will be in AdS until otherwise specified. Furthermore, we will focus on the semiclassical limit where $G_{N} \rightarrow 0$ in our applications of the generalized entanglement wedges.

\section{Preliminaries}\label{sec:prelim}

The aim of this paper is to bring together two frameworks which highlight how spacetime encodes nontrivial aspects of quantum information: the \textit{connected wedge theorem }\cite{May:2019odp,May:2021nrl, Caminiti:2024ctd}, and the recent proposal \cite{Bousso:2022hlz,Bousso:2023sya} for generalizing the notion of an \textit{entanglement wedge} to bulk regions. In this section we will review the salient definitions and results from these constructions.

\subsection{The Connected Wedge Theorem}\label{sec:cwt}

In the semiclassical regime of AdS/CFT, the bulk spacetime is emergent from the boundary state.  Different CFT states correspond to different bulk geometries, and thus to different causal structures. As a result, bulk causal structure can be very rich, and can be used to probe the entanglement structure of the CFT state. This is seen concretely in the connected wedge theorem, which addresses situations with a \textit{causal discrepancy}, which occurs when a classical scattering process is permitted by bulk causal structure, but forbidden in the boundary.  This arises naturally in the context of quantum tasks, which are essentially quantum computations with inputs and outputs at specified spacetime locations.  

There are generally two approaches for completing a quantum task: local and non-local. In the local strategy the agents bring the inputs together to some scattering point where the quantum channel is applied and then deliver the outputs to their destinations. Alternatively, the task can be completed with a non-local strategy, in which the agents make use of entanglement resources to complete the task without relying on a scattering point. In fact, it can be shown that the agents can generally replace the local strategy by entanglement \cite{Kent2006TaggingSystemsGranted,Kent_2011, Buhrman_2014, dolev2022constrainingdoabilityrelativisticquantum}. In the holographic context, the only possible entanglement resource comes from the CFT state. Hence, we expect that when a causal discrepancy occurs, there must exist sufficient entanglement resources on the boundary to replace the local protocol. This is exactly what the connected wedge theorem tells us.

\begin{figure}
    \centering
    \includegraphics[width=1\linewidth]{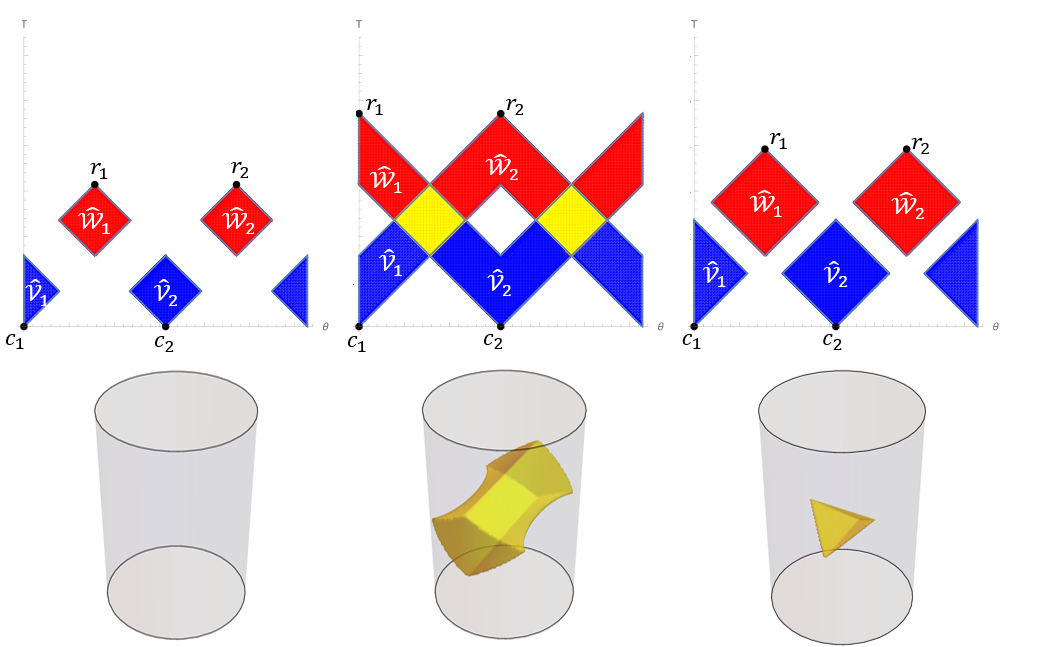}
    \caption{Three possible scattering configurations in pure global $\text{AdS}_{3}$. The first row shows the boundary geometry unwrapped (so the left and right end are identified) with $\hatit{V}_{i}$ shown in blue, $\hatit{W}_{i}$ shown in red and $s_{\partial, pts}$ shown in yellow. The second row shows the bulk scattering region. \textit{Left}: A configuration in which both the bulk and boundary scattering region is empty. \textit{Center}: A configuration in which  both the bulk and boundary scattering region are non-empty. The bulk scattering region extends all the way the the boundary. \textit{Right}: A bulk-only scattering configuration.}
    \label{fig: example scattering configurations}
\end{figure}

Let us consider a quantum tasks with two input points  $c_{1}$ and $c_{2}$ on the asymptotic boundary, and two output points $r_{1}$ and $r_{2}$ also on the asymptotic boundary.  Local scattering is possible on the boundary if the boundary scattering region:
\begin{equation}
    s_{\partial, pts} = \hat{J}^{+}(c_{1}) \cap \hat{J}^{+}(c_{2}) \cap \hat{J}^{-}(r_{1}) \cap \hat{J}^{-}(r_{2}) 
\end{equation}
is non-empty. Analogously, local scattering is possible in the bulk when the $2\rightarrow 2$ region:
\begin{equation}
    s_{pts} = J^{+}(c_{1}) \cap J^{+}(c_{2}) \cap J^{-}(r_{1}) \cap J^{-}(r_{2}) 
\end{equation}
is non-empty. The interesting feature of $\text{AdS}_{3}$ is that it is possible to choose the scattering configuration $\{c_{1}, c_{2}, r_{1}, r_{2}\}$ such that we have a causal discrepancy, i.e.  $s_{pts} \neq \emptyset$, but $s_{\partial, pts} = \emptyset$. This is called a \textit{bulk-only scattering configuration}, depicted in the right panel of Figure \ref{fig: example scattering configurations}.

We may define the \textit{decision regions} $\hatit{V}_{i}$ and $\hatit{W}_{i}$ for $i = 1,2$:
\begin{align}
    \hatit{V}_{i} := \hat{J}^{+}(c_{i}) \cap \hat{J}^{-}(r_{1}) \cap \hat{J}^{-}(r_{2}) \\
    \hatit{W}_{i} := \hat{J}^{+}(c_{1}) \cap \hat{J}^{+}(c_{2}) \cap \hat{J}^{-}(r_{i}).
\end{align}
$\hatit{V}_{i}$ corresponds to the region for which a signal from $c_{i}$ can reach both output points (and similar for $\hatit{W}_{i}$), namely the boundary restriction of $1\rightarrow 2$ (or $2\rightarrow 1$) scattering region. The boundary geometry and bulk scattering region are depicted in Figure \ref{fig: example scattering configurations}.

\begin{thm}[\textbf{Points-Based Connected Wedge Theorem}]
    Let $\{c_{1}, c_{2}, r_{1}, r_{2}\}$ be a bulk-only scattering configuration on the boundary of an asymptotically AdS spacetime with a holographic dual such that the bulk scattering region is nonempty $s_{pts} \neq \emptyset$. Then for decision regions $\hatit{V}_{1}$ and $\hatit{V}_{2}$ defined as above,
    \begin{equation}
        I(\hatit{V}_{1}; \hatit{V}_{2}) \geq O(1/G_{N}).
    \end{equation}
\end{thm}

 We may use the HRRT prescription to reformulate the mutual information in terms of entanglement wedges. $I(\hatit{V}_{1}; \hatit{V}_{2}) \geq O(1/G_{N})$ means that $\E{\hatit{V}_{1}} \cup \E{\hatit{V}_{2}} \neq \E{\hatit{V}_{1} \cup {\hatit{V}_{2}}}$ and  implies that the entanglement wedge $\ew{\hatit{V}_{1} \cup {\hatit{V}_{2}}}$ is connected. Generally, the converse of the CWT does not hold: $I(\hatit{V}_{1}; \hatit{V}_{2}) \geq O(1/G_{N})$ does not imply a non-empty bulk scattering region. Counterexamples are provided in \cite{May:2019odp,Caminiti:2024ctd}.

For what follows it will also be useful to include a stronger version. Instead of considering input and output points $\{c_{1}, c_{2}, r_{1}, r_{2}\}$, we now consider input and output \textit{regions} $\{\hatit{V}_{1}, \hatit{V}_{2}, \hatit{W}_{1}, \hatit{W}_{2}\}$ \cite{May:2021nrl} defined as above. Since these regions correspond to bulk regions $\{\ew{\hatit{V}_{1}}$, $\ew{\hatit{V}_{2}}$, $\ew{\hatit{W}_{1}}$, $\ew{\hatit{W}_{2}}\}$, this allows us to consider bulk scattering processes where the input and output points are not located at asymptotic infinity, but rather lie within the bulk, inside the entanglement wedge of the boundary decision regions. We may define a larger scattering region, called the regions-based scattering region $s_{reg}$ by:
\begin{equation}
    s_{reg} = J^{+}(\ew{\hatit{V}_{1}}) \cap J^{+}(\ew{\hatit{V}_{2}}) \cap J^{-}(\ew{\hatit{W}_{1}}) \cap J^{-}(\ew{\hatit{W}_{2}}).
\end{equation}

\begin{thm}[\textbf{Regions-Based Connected Wedge Theorem}]
   Let $\{c_{1}$, $c_{2}$, $r_{1}$,$ r_{2}\}$ be a bulk only scattering configuration on the boundary of an asymptotically AdS spacetime with a holographic dual with decision regions $\{\hatit{V}_{1}$, $\hatit{V}_{2}$, $\hatit{W}_{1}$, $\hatit{W}_{2}\}$. Then if the regions-based scattering region is non-empty ($s_{reg} \neq \emptyset$),
    \begin{equation}
        I(\hatit{V}_{1}; \hatit{V}_{2}) \geq O(1/G_{N})
    \end{equation}
    i.e. the entanglement wedge $\ew{\hatit{V}_{1} \cup \hatit{V}_{2}}$ is connected.
\end{thm}
Note that the region-based CWT is stronger than the point-based one. The points $c_{i}$ are the past boundary points of the decision regions $\hatit{V}_{i}$ and the points $r_{i}$ are the future boundary points of $\hatit{W}_{i}$ so $s_{pts} \subset s_{reg}$.\footnote{In the special case of pure global $\text{AdS}_{3}$, the fact that the causal wedge equals the entanglement wedge, implies that $s_{pts} = s_{reg}$.} Therefore, it is sufficient to prove the regions-based CWT. Note that the converse of the theorem does not hold even when restricted to the regions bulk scattering region: $I(\hatit{V}_{1}, \hatit{V}_{2}) \geq O(1/G_{N})$ does not imply $s_{reg} \neq \emptyset$.

\begin{figure}
    \centering
    \includegraphics[width=0.6\linewidth]{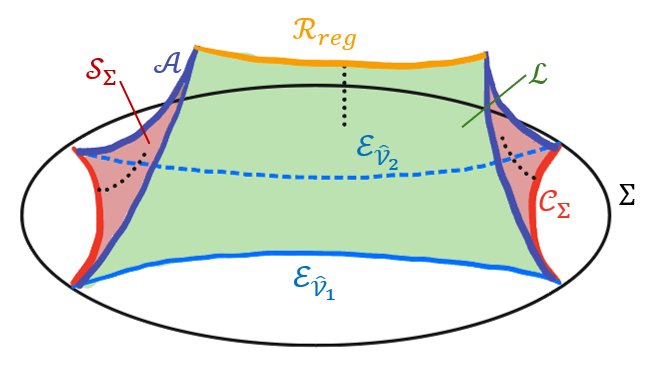}
    \caption{The lift $\mathcal{L}$ (green) and slope $\mathcal{S}_{\Sigma}$ (dark red) and their boundary, composed of: the ridge $\mathcal{R}_{reg}$ (orange), the extremal surface $\E{\hatit{V}_{1}} \cup \E{\hatit{V}_{2}}$ (blue), the contradiction surface $\mathcal{C}_{\Sigma}$ (red), the seams $\mathcal{A}$ between the lift and slope (dark blue), and possible cusps $\mathcal{B}_{\mathcal{L}}$ and $\mathcal{B}_{\mathcal{S}_{\Sigma}}$ depicted as dotted lines.}
    \label{fig: null membrane}
\end{figure}

The gravitational proof of the CWT is given in \cite{May:2019odp,May:2021nrl}. While we will not review the proof in detail here, it is worth emphasizing that the basic elements of the proof since we will need similar methods in section~\ref{sec:scatreg}. The main ingredient is the focusing theorem:
\begin{thm}[Focusing Theorem]
\label{thm: focusing thm}
    Let $M$ be a spacetime satisfying the null energy condition. Then for any hypersurface orthogonal affinely parametrized null congruence, the expansion is non-increasing:
    \begin{equation}
        \frac{d \theta}{d \lambda} \leq 0. 
    \end{equation}
\end{thm}
\noindent This follows from the Raychaudhuri equation \cite{Raychaudhuri:1953yv} and the null energy condition. For the CWT, we are specifically interested in lightsheets normal to extremal surfaces, that is null surfaces $\mathcal{N}^{\pm} = \partial J^{\pm}(\mathcal{E})$ where $\mathcal{E}$ is an extremal surface. Since $\theta = 0$ on the extremal surface, the focusing theorem implies that $\theta \leq 0$ everywhere on the null sheet. We can also apply Stokes' theorem to conclude: 
\begin{thm}[Area Theorem~\cite{May:2019odp}] 
    Let $\mathcal{N}$ be a codimension-1 null surface in a $(d+1)$ dimensional spacetime. Assume that $\partial \mathcal{N}$ consists of two spacelike surfaces $\Sigma_{1}$ and $\Sigma_{2}$ such that $\Sigma_{2}$ is in the future of $\Sigma_{1}$. Suppose in addition that $\mathcal{N}$ is generated by null geodesics with affine parameter $\lambda$. Then
    \begin{equation}
        \area{\Sigma_{2}} - \area{\Sigma_{1}} = \int_{\mathcal{N}} \theta \mathbf{\epsilon}_{a_{1}...a_{d}}
    \end{equation}
    where $\mathbf{\epsilon}_{a_{1}...a_{d}}$ denotes the volume form on $\mathcal{N}$ and $\theta$ denotes the usual classical expansion $\theta = h^{ab} \nabla_{a} k_{b}$ with respect to the generators of the geodesics $k^{a}$ .
\end{thm}
\noindent Applying this to judiciously chosen null congruences, one can then establish not only the above theorems, but also the lower bound
\begin{equation}
\label{eq: CWT result}
    I(\hatit{V}_{1}; \hatit{V}_{2}) \geq \frac{1}{2 G_{N}}\area{\mathcal{R}_{reg}}  
\end{equation}
where the ridge ${\cal R}_{reg}$ is given by 
\begin{equation}
\label{eq: ridge def}
    \mathcal{R}_{reg} = \partial J^{+}(\ew{\hatit{V}_{1}}) \cap \partial J^{+}(\ew{\hatit{V}_{2}}) \cap J^{-}(\ew{\hatit{W}_{1}})  \cap J^{-}(\ew{\hatit{W}_{2}}) .
\end{equation}
The focusing set up is illustrated in figure~\ref{fig: null membrane}. In brief, we are focusing up from the candidate RT surface ${\cal E}_{\V_1}\cup {\cal E}_{\V_2} $, to a ridge ${\cal R}_{reg}$, and back along a slope ${\cal S}_{E}$ set by the past lightsheets from the  $e_W(\W_i)$, to a contradiction surface $\C_\Sigma$ which must be longer in length than the RT surface with the same end points by the maximin property of the Cauchy slice $\Sigma$. The signs in the focusing proof are such that ${\cal E}_{\V_1}\cup {\cal E}_{\V_2} $ is thus established to be longer than another candidate for ${\cal E}_{\V_1 \cup \V_2} $, and thus the wedge is connected. We know that these surfaces have the topology illustrated in figure~\ref{fig: null membrane} due to the fact that the ridge ${\cal R}_{reg}$ forms the past boundary of the regions-based bulk scattering region $s_{reg}$. Therefore, since we have assumed that $s_{reg} \neq 0$, we indeed see the ridge is non-empty.

\subsection{Generalized Entanglement Wedges}
Let us now turn to the motivation and proposal for generalized entanglement wedges given in \cite{Bousso:2022hlz,Bousso:2023sya} (also colloquially referred to as the BP proposal). The HRRT~\cite{Ryu:2006bv,Hubeny:2007xt} result can be derived from the gravitational path integral \cite{Lewkowycz:2013nqa}, suggesting that it should be extendable to more general gravitational systems, not just asymptotically AdS spacetimes. However, one of the challenges in developing the entanglement wedge prescription in non-AdS spacetimes is the absence of extremal surfaces anchored on the boundary. For instance, in Minkowski space, if we choose two generic points on the conformal boundary, there is no bulk geodesic connecting them. In \cite{Bousso:2022hlz,Bousso:2023sya}, Bousso and Penington suggest that instead of tying the entanglement wedge prescription to the conformal boundary, we should instead be assigning an entanglement wedge to \textit{bulk} regions. More specifically, given a bulk region $a$, we want to define a larger bulk region $e(a)$ which we can interpret as the entanglement wedge of $a$. Then the boundary entanglement wedge proposal would be understood as a limiting case of the more general bulk construction.

To motivate this construction, let us begin by considering more carefully what we mean when we say that the entanglement wedge $\ew{\hatit{A}}$ is dual to the boundary subregion $\hatit{A}$. We would like $\ew{\hatit{A}}$ to be the region which contains all the information (and only the information) contained in $\hatit{A}$. In terms of bulk and boundary operators, this means that
\begin{enumerate}
    \item All bulk operators in $\ew{\hatit{A}}$ can be reconstructed from boundary operators in $\hatit{A}$.
    \item No bulk operators outside $\ew{\hatit{A}}$ can be reconstructed from boundary operators in $\hatit{A}$.
\end{enumerate}
The quantum extremal surface (QES) prescription tells us that the entanglement entropy of a boundary subregion $\hatit{A}$ is given by:
\begin{equation}
    S(\hatit{A}) = S_{gen}(\E{\hatit{A}})
\end{equation}
where $\E{\hatit{A}}$ is the surface homologous to $\hat{A}$ which extremizes the generalized entropy: 
\begin{equation}
    S_{gen}(\mathcal{X}) := \frac{\area{\mathcal{X}}}{4 G_{N}} + S(a_{\mathcal{X}}) + \text{counterterms} .
\end{equation}
Here $S(\hatit{A})$ is the von Neumann entropy of the boundary region $\hatit{A}$, while $S(a_{\mathcal{X}})$ is the von Neumann entropy of the bulk fields. When the QES prescription holds, it can be shown that the bulk matter inside the associated entanglement wedge is encoded in the boundary state in $\hatit{A}$ \cite{Dong:2016eik}, in the sense that bulk operators in $\ew{\hatit{A}}$ can be reconstructed from operators in $\hatit{A}$.

The bulk entropy term would nominally seem like a small correction in the semiclassical regime, since the area term is $O(1/G_{N})$. However, it can be important in the context of black hole evaporation~\cite{Penington:2019npb, Almheiri:2019psf, Almheiri:2019hni}. Indeed the entanglement island story motivates more generally analyzing regimes where bulk entropy is significant. In \cite{Akers:2020pmf, Akers:2023fqr} it was found that the QES prescription requires modification in cases where extremal surface candidates have $O(1/G_{N})$ bulk entropy between them. In such cases, they find that it is not possible to define a single entanglement wedge $\ew{\hatit{A}}$ which satisfies both properties 1 and 2. Instead, they define two distinct entanglement wedges, $\textit{max-}\ew{\hatit{A}}$ and $\textit{min-}\ew{\hatit{A}}$, each satisfying one of the two properties individually:
\begin{enumerate}
    \item $\textit{max-}\ew{\hatit{A}}$ is the largest bulk region within which all operators can be reconstructed from boundary operators in $\hatit{A}$.
    \item $\textit{min-}\ew{\hatit{A}}$ is the smallest bulk region outside which no operators can be reconstructed from boundary operators in $\hatit{A}$.
\end{enumerate}
Instead of defining a von Neumann entropy, the min/max entanglement wedges are defined in terms of a quantity called the \textit{smooth conditional min/max entropy} which arises from the study of one-shot Shannon theory and is closely related to the quantum information task of \textit{quantum state merging} \cite{berta2009singleshotquantumstatemerging, Horodecki_2006}.  When $\textit{max-}\ew{\hatit{A}} = \textit{min-}\ew{\hatit{A}}$, they coincide with the QES prescription $\ew{\hatit{A}}$ for the entanglement wedge. This occurs for a class of states known as ``compressible'' states for which the smooth conditional min/max entropies are equal to the von Neumann entropy. 

Moreover, the fact that entanglement islands are necessarily disconnected from the boundary signals that the notion of an entanglement wedge need not be tied to the conformal boundary.
Both of these motivate the construction of a {\it generalized entanglement wedge}. A definition, for the simpler case of time-independent settings was given in \cite{Bousso:2022hlz}. This was generalized to time-dependent contexts in \cite{Bousso:2023sya} and refined in~\cite{Bousso:2024iry,Bousso:2025fgg}. Importantly, in order to establish properties like no-cloning, nesting, and strong-subadditivity they needed to introduce the notion of a min and max entanglement wedge for a bulk region $a$: $e_{min}(a)$ and $e_{max}(a)$.  These coincide with $\textit{min-}\ew{\hatit{A}}$ and $\textit{max-}\ew{\hatit{A}}$, in the limit where $a$ becomes an asymptotic region. Crucially, unlike $\textit{min-}\ew{\hatit{A}}$ and $\textit{max-}\ew{\hatit{A}}$, $e_{min}(a)$ and $e_{max}(a)$ can differ at even the classical level in time-dependent geometries. This seems to suggest that classical gravity encodes information about one-shot Shannon entropy and quantum state merging, further illustrating how quantum information is embedded in the structure of gravity.

We will now proceed to define the generalized entanglement wedges of bulk regions \cite{Bousso:2024iry, Bousso:2025joj, Bousso:2025fgg}. Before doing so, we will need some causal structure definitions that appear in those proposals.
\begin{defn}
    The \textit{spacelike complement} of a set $s \subset M$ is:
    \begin{align}
        s' = M - \left[\text{cl}\left(J^{+}(s)\right) \cup \text{cl}\left(J^{-}(s)\right)\right] .
    \end{align}
\end{defn}
\begin{defn}
    A \textit{wedge} is a set $a \subset M$ such that
    \begin{align}
        a'' = a .
    \end{align}
\end{defn}
\noindent If $a$ is a wedge, then its spacelike complement $a'$ is also a wedge. Additionally, the intersection of two wedges $a$ and $b$ is a wedge. However, the union $a \cup b$ is not necessarily a wedge.
\begin{defn}
    Given two wedges $a$, $b$ the \textit{wedge union} given by
    \begin{align}
        a \doublecup b = (a' \cap b')'
    \end{align}
    is a wedge and contains $a \cup b$.
\end{defn}

\begin{defn}
    Given a wedge $a$, let $C$ be the set of points that lie on timelike curves that are entirely contained in $a$ and have infinite proper time duration to the past and to the future. The \textit{causal wedge in} $a$ is \cite{Bousso:2025fgg}:
    \begin{equation}
        a_{C} = C''.
    \end{equation}
\end{defn}
\noindent Note that the causal wedge in $a$ is a wedge, and is contained in $a$.
\begin{defn}
    The \textit{fundamental complement}  $\tilde{a}$ of a wedge $a$ is the causal wedge of its spacelike complement $a'$ \cite{Bousso:2025fgg}:
    \begin{equation}
        \tilde{a} = a'_{C} .
    \end{equation}
\end{defn}
\begin{defn}
    The \textit{edge} $\eth a$ of a wedge $a$ is given by
    \begin{align}
        \eth a = \partial a \cap \partial a' .
    \end{align}
\end{defn}
\noindent  We then note that the edge of the intersection of two wedges $a$ and $b$ is given by:
\begin{equation}
    \eth (a \cap b )  = \text{cl}\left[(\eth a \cap b) \sqcup (H^{+}(a) \cap H^{-}(b)) \sqcup \{ a \leftrightarrow b\}\right] .
    \label{eq: edge of intersection}
\end{equation}
\begin{defn}
    The \textit{future/past expansion} $\theta^{\pm}$ of a wedge $a$ at some $p \in \eth a$, is the expansion of the outward future/past directed null congruence orthogonal to $\eth a$ at $p$.
\end{defn}

\begin{defn}
    At $p \in \eth a$, a wedge $a$ is called 
\begin{itemize}
    \item \textit{normal} if $\theta^{+} \geq 0$ and $\theta^{-} \geq 0$
    \item \textit{antinormal} if $\theta^{+} \leq 0$ and $\theta^{-} \leq 0$
    \item \textit{trapped} if $\theta^{+} \leq 0$ and $\theta^{-} \geq 0$
    \item \textit{antitrapped} if $\theta^{+} \geq 0$ and $\theta^{-} \leq 0$
    \item \textit{extremal} if $\theta^{+} = 0$ and $\theta^{-} = 0$
\end{itemize}
\end{defn}

\begin{defn}
    The \textit{generalized entropy} $S_{gen}(a)$ of a wedge $a$ is:
    \begin{align}
        S_{gen}(a) = \frac{\text{Area}(\eth a)}{4 G_{N} } + S(a)
    \end{align}
where $S(a)$ is the von Neumann entropy of the reduced quantum state of the matter fields on any Cauchy slice of $a$.
\end{defn}
For the purposes of this paper we will be working in the classical limit $G_N\rightarrow 0$. Therefore, we will ignore $S(a)$ and keep only the area term.\footnote{When we include the quantum state of the bulk, the expansions should be replaced with the \textit{quantum expansion}, the shape derivative of the generalized entropy under outward deformations of $a$:
\begin{equation}
    \Theta^{\pm} = \theta^{\pm} + 4 G_{N} \frac{\delta S(a)}{\delta \lambda^{\pm}}
\end{equation}
where $\lambda^{\pm}$ is an affine parameter for the future/past null congruence orthogonal to $\eth a$. The focusing theorem (Theorem \ref{thm: focusing thm}) is then replaced with the \textit{quantum focusing conjecture} \cite{Bousso:2015mna, Leichenauer:2017bmc}. Additionally, to be consistent with the refinements \cite{Bousso:2024iry,Bousso:2025joj,Bousso:2025fgg} Definitions~\ref{def:max} and \ref{def:min} would be modified to include smooth conditional max- and min-entropies of the matter fields in \eqref{eq: sgen f < sgen h} and \eqref{eq: sgen g < sgen h}. In the classical limit where $S_{gen}(a) = \area{\eth a}/4 G_{N}$, these definitions are equivalent since $H_{min/max, gen}(f|h) \rightarrow (1/4 G_{N})\left(\area{\eth f} - \area{\eth h}\right)$.
} We are now ready to define the generalized entanglement wedge of a gravitating region.
\label{sec: ew proposal}
\begin{defn}[Max-Entanglement Wedge]\label{def:max}
    Given a wedge $a$, the max-entanglement wedge $e_{max}(a)$ is the wedge union
    \begin{align}
        e_{max}(a) = \doublecup_{f \in F(a)} f
    \end{align}
    of all wedges $f \in F(a)$ where $F(a)$ is the set of wedges satisfying:
    \begin{enumerate}
        \item $a \subset f \subset \tilde{a}'$ 
        \item $f$ is antinormal for all $p \in \eth f - \eth a$
        \item $f$ admits a Cauchy slice $\Sigma$ with $\eth a \subset \Sigma$ such that for any other wedge $h \neq f$ satisfying $a \subset h$, $\eth h \subset \Sigma$, and $\eth h - \eth f$ is compact, we have:
            \begin{align}
                S_{gen}(f) < S_{gen}(h) .
                \label{eq: sgen f < sgen h}
            \end{align}
    \end{enumerate}
\end{defn}

\begin{defn}[Min-Entanglement Wedge]\label{def:min}
    Given a wedge $a$, the min-entanglement wedge $e_{min}(a)$ is the intersection:
    \begin{equation}\label{eq:intmin2}
        e_{min}(a) = \cap_{g\in G(a)} g
    \end{equation}
    of all wedges $g \in G(a)$, where $G(a)$ is the set of wedges satisfying:
    \begin{enumerate}
        \item $a \subset g \subset \tilde{a}'$
        \item $g$ is normal
        \item $g' \cap  \tilde{a}'$ admits a Cauchy slice $\Sigma'$ such that for any wedge $h \neq g$ such that $g \subset h \subset \tilde{a}'$, $\eth h \subset \Sigma'$, and $\eth h - \eth g$ is compact we have:
        \begin{equation}
            S_{gen}(g) < S_{gen}(h) .
            \label{eq: sgen g < sgen h}
        \end{equation}
    \end{enumerate}
\end{defn}

\noindent Note that $e_{max}(a) \in F(a)$ and $e_{min}(a) \in G(a)$ \cite{Bousso:2023sya}.  An immediate consequence is that for any wedge $a$, we have:
\begin{equation}
    S_{gen}(e_{max}(a)) \leq S_{gen}(a) .
    \label{eq: S(emax) < S(a)}
\end{equation}
This follows by setting $f = e_{max}(a)$ in property 3 of $F(a)$, and choosing $h = a$.
To support their interpretation as min and max entanglement wedges, the $e_{min}$ and $e_{max}$ are shown to satisfy several properties: 
\begin{thm}
\label{thm: properties of emin and emax}
\textit{Properties of $e_{min}(a)$ and $e_{max}(a)$ \cite{Bousso:2023sya, Bousso:2025fgg}} 
\begin{itemize}
    \item \textit{Inclusion: }$e_{max}(a) \subset e_{min}(a)$. 
    \item \textit{Nesting: } For wedges $a$ and $b$ such that $a \subset b$,
    \begin{equation}
        e_{min}(a) \subset e_{min}(b) .
    \end{equation}
    Additionally, if $\eth e_{min}(b)  - \eth e_{min}(a)$ is compact,
    \begin{equation} \label{eq:nestmin}
        S_{gen}(e_{min}(a)) \leq S_{gen}(e_{min}(b)).
    \end{equation}
    \item \textit{Complementarity: } $e_{min}(a)' = e_{max}(\tilde{a})|_{a'}$.\footnote{ Here, $e_{max}(\tilde{a})|_{a'}$ denotes the max-entanglement as computed in the $a'$ spacetime.}
\end{itemize}   
\end{thm}
\noindent Additionally, when  $e_{min}(a) = e_{max}(a) := e(a)$, (and for $b$, $c$, and their wedge unions), i.e. when we can assign a single entanglement wedge to these regions, the generalized entropy of these regions satisfy \textit{strong subadditivity} (SSA)
    \begin{equation}
        S_{gen}(e(a \doublecup b \doublecup c)) + S_{gen}(e(b)) \leq S_{gen}(e(a \doublecup b) ) + S_{gen}(e(b \doublecup c)) .
\end{equation}
Finally, we note one additional property of $e_{min}(a)$ shown in \cite{Bousso:2023sya} which will be useful to us in the next section.
\begin{thm}\label{minedge}
    $e_{min}(a)$ is marginal or extremal at 
    points $p \in \eth e_{min}(a) - \eth a$. Specifically,
    \begin{itemize}
        \item $e_{min}(a)$ is marginally anti-trapped  $\theta^{-} = 0$ at $p \in \eth e_{min}(a) \cap H^{+}(a')$.
        \item $e_{min}(a)$ is marginally trapped   $\theta^{+} = 0$ at $p \in \eth e_{min}(a) \cap H^{-}(a')$.
        \item $e_{min}(a)$ is extremal $\theta^{+} = \theta^{-} = 0$ at $p \in \eth e_{min}(a) \cap a'$.
    \end{itemize}
\end{thm}

\section{Bounding the Mutual Information with Bulk Entropies}\label{sec:scatreg}

We now apply the generalized entanglement proposal to the bulk scattering regions defined in section \ref{sec:cwt}. We will utilize the same type of focusing arguments used in the proof of the connected wedge theorem to relate the entropy of the bulk region to its past boundary. Additionally, we can relate the entropy of the scattering region to the entropy of their min/max entanglement wedges using the properties of $e_{min}$ and $e_{max}$. The main results of this section will be
\begin{equation}\label{bound}
    S_{gen}(e_{max}(s_{pts}'')) \leq I(\hatit{V}_{1}; \hatit{V}_{2}) \leq S_{gen}(e_{min}(s_{ent})) 
\end{equation}
and similarly with $s_{pts}$ replaced with $s_{reg}$, as well as
\begin{equation}\label{nest}
    S_{gen}(e_{min}(s_{pts}'')) \le S_{gen}(e_{min}(s_{reg}'')) \le S_{gen}(e_{min}(s_{ent})) .
\end{equation}
Here $s_{pts}$ and $s_{reg}$ were defined in section~\ref{sec:cwt}, while $s_{ent}$ is a larger scattering region we will call the {\it entanglement scattering region}
\begin{equation} \label{s-ent}
    s_{ent} := \ew{\hatit{V}_{1} \cup \hatit{V}_{2}} \cap \ew{\hatit{W}_{1} \cup \hatit{W}_{2}}
\end{equation}
defined in~\cite{Leutheusser:2024yvf} in connection to a proposed generalized CWT with converse.\footnote{This is explored in~\cite{Lima:2025dtj}.} The lower bound in~\eqref{bound} will be proven in section~\ref{sec:lower} and the upper bound in section~\ref{sec:upper}. Meanwhile~\eqref{nest} follows immediately from the nesting property~\eqref{eq:nestmin} given $s_{pts}\subset s_{reg}\subset s_{ent}$ as established in~\cite{May:2021nrl,Leutheusser:2024yvf}.\footnote{As we will see below $s_{ent}$ is already a wedge while the double complements are the smallest wedges containing $s_{pts/reg}$, and so the nesting  $s_{pts}\subset s_{reg}\subset s_{ent}$ implies $s_{pts}''\subset s_{reg}''\subset s_{ent}$. 
}

Moreover, since the entanglement wedge of a boundary region can be replaced with a bulk definition, we can view~\eqref{bound} as a fully bulk statement of the connected wedge theorem. More precisely, local CFT operators located in a boundary subregion $\hatit{A}$ are dual to bulk operators near $\hatit{A}$ \cite{banks1998adsdynamicsconformalfield, Hamilton_2006}. Since local CFT operators generate the entire CFT algebra, we expect that the algebra of the operators in the near-boundary bulk region $a$ should generate all the operators inside the entanglement wedge $\ew{\hatit{A}}$. Therefore, we can view the entanglement wedge $\ew{\hatit{A}}$ as really being the entanglement wedge $e(a)$ of the bulk region near the boundary. Since the $\textit{max-}\ew{\hat \A}$ and $\textit{min-}\ew{\hat \A}$ of our boundary regions will be equal in our semiclassical context, and since \cite{Bousso:2025fgg} established\footnote{Technically their definition for the causal wedge is $\mathrm{CW}({\hat \A}) \equiv\left(\hat \A_{\tilde{M}}^{\prime} \cap M\right)^{\prime}$. This will be the same as ours up to subtleties with closure that will not change the conclusions of our proof. 
}
\be\label{eq:ecw}
\textit{max-}\ew{\hat \A}=e_{max}(\cw{\hat \A}),~\textit{min-}\ew{\hat \A}=e_{min}(\cw{\hat A}),
\ee
we can recognize
\be\label{mibulk}
I(\hatit{V}_{1}; \hatit{V}_{2})=S_{gen}(e(\cw{\V_1})+S_{gen}(e(\cw{\V_2}))-S_{gen}(e(\cw{\V_1}\Cup \cw{\V_2})) 
\ee
where $e(a)=e_{min}(a)=e_{max}(a)$. Now the theorems of Bousso et al. guarantee~\eqref{mibulk} satisfies SSA, and phrased in this way we indeed have a bulk statement of the entropy bounds. There are other equivalent ways for writing~\eqref{mibulk}, but this version in terms of bulk causal wedges will be closest to what we will use in section~\ref{sec:bulkcwt}. Indeed, as we will elaborate on more there, we see that we can rephrase the connected wedge theorem as the statement that bulk scattering implies $e(\cw{\V_1}\Cup \cw{\V_2})$ is connected. 

\subsection{Lower Bounds from Bulk Scattering Regions} \label{sec:lower}
\begin{figure}
    \centering
    \includegraphics[width=0.5\linewidth]{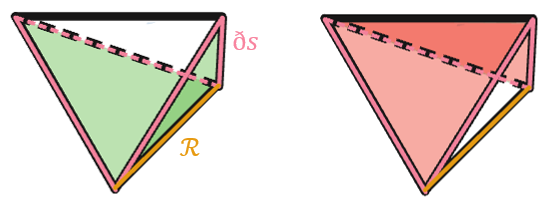}
    \caption{The scattering region for $s_{pts}$ or $s_{reg}$ has a tetrahedral-like shape as depicted above. The past Cauchy horizon $H^{-}(s)$ shaded in green in the left image is generated by the future-moving causal horizons of $c_{1}$ and $c_{2}$ for $s_{pts}$ or the future-moving null sheet of $\E{\hatit{V}_{1}}$ and $\E{\hatit{V}_{2}}$ for $s_{reg}$. Consequently, at the edge $\eth s$, the future expansion is negative. The right image shows the future boundary shaded in red. This is generated by the past-moving causal horizon/lightsheet, and consequently at the edge $\eth s$, the past expansion is negative. 
    }
    \label{fig: spts and sreg}
\end{figure}

We will now establish the lower bounds in~\eqref{bound}. The first step is to create regions that we know are wedges from $s_{pts}$ and $s_{reg}$. We know that the boundary of these regions consist of components of the lightsheets $\partial J^{+}(c_{i})$, $\partial J^{-}(r_{j}) $  and $\partial J^{+}(\ew{\hatit{V}_{i}})$, $\partial J^{-}(\ew{\hatit{W}_{j}}) $, respectively.  We thus see that the segments\footnote{This is a slight abuse of notation to use $\eth s_{pts/reg}$ to denote this edge when $s_{pts/reg}$ is not necessarily a wedge. Also note that while we can write $s_{pts}$ and $s_{reg}$ as intersections of the form $J^{+}(c_{i}) \cap J^{-}(r_{j})$ or $J^{+}(\ew{\hatit{V}_{i}}) \cap J^{-}(\ew{\hatit{W}_{j}})$, respectively, it is not obvious that these components are themselves wedges, and so we cannot immediately apply~\eqref{eq: edge of intersection}.}
\begin{align}\label{edges}
    \eth s_{pts} &= \bigcup_{i,j} \left(\partial J^{+}(c_{i}) \cap \partial J^{-}(r_{j}) \cap s_{pts}\right) \\
    \eth s_{reg} &= \bigcup_{i,j} \left(\partial J^{+}(\ew{\hatit{V}_{i}}) \cap \partial J^{-}(\ew{\hatit{W}_{j}}) \cap s_{reg}\right)
\end{align}
form closed spacelike curves. From nesting of past/future lightcones we can also conclude that these ``edges'' are spacelike from all points in the respective interiors of $s_{pts/reg}$, and will form the edge of the spacelike complement $s_{pts/reg}'$ which is a wedge by Corollary 48 of~\cite{Bousso:2025fgg}. We can thus promote~\eqref{edges} to a statement about $s_{pts/reg}''$ which is guaranteed to be a wedge.\footnote{Note that for the configurations we are considering we expect $s_{pts}$ and $s_{reg}$ to be a single connected component, and the topology illustrated in figure~\ref{fig: spts and sreg}. In particular the future (past) Cauchy horizon can also be written as a segment of the boundary of the past (future) of the outgoing (incoming) ridge, which will be relevant in section~\ref{sec:bulkcwt}.  \label{ft:sptstop}
}

Note that $s_{pts/reg}\subseteq s_{pts/reg}''$ can differ in situations where we have cusps. While this changes $H^\pm(s_{pts/reg}'')$, our focusing proofs in what follows will still use the lightsheets from $\{c_i,r_j\}$ and $\{v_i,w_j\}$, respectively, to make statements about the edge of the wedge, which is the same as~\eqref{edges}.  
Away from such cusps, the wedges $s_{pts}''$ and $s_{reg}''$ are antinormal, because the future/past Cauchy horizon is generated by a past/future moving lightsheet from either a point on the boundary (points based), or an extremal surface (regions based). We call the past boundary of the scattering regions $s_{pts}$ and $s_{reg}$ the ``ridge'' $\mathcal{R}_{pts}$ and $\mathcal{R}_{reg}$ respectively, given by:
\begin{align}
    \mathcal{R}_{pts} &=  \partial J^{+}(c_{1}) \cap \partial J^{+}(c_{2}) \cap J^{-}(r_{1})  \cap J^{-}(r_{2}) \\
    \mathcal{R}_{reg} &= \partial J^{+}(\ew{\hatit{V}_{1}}) \cap \partial J^{+}(\ew{\hatit{V}_{2}}) \cap J^{-}(\ew{\hatit{W}_{1}})  \cap J^{-}(\ew{\hatit{W}_{2}}) .
\end{align}
This is illustrated in Figure \ref{fig: spts and sreg}.

\begin{thm}
\label{thm: sgen(spts) leq ridge}
    $S_{gen}(s_{pts}'') \leq \frac{1}{2 G_{N}} \area{\mathcal{R}_{pts}}$ 
\end{thm}
\noindent To show this, we will again use the fact that the faces of the scattering region are generated by \textit{causal horizons}, that is the boundary of the future/past of a point at infinity. Similar to the case of lightsheets of extremal surfaces, the focusing theorem implies that expansion of a future/past causal horizon $\partial J^{\pm}$ is non-positive with respect to future/past directed generators. Namely, not only do we have the antinormal property at the edge of the wedge, we know how to focus this edge along the faces of the scattering region.

Consider the null surface $\mathcal{N}$ defined by:
\begin{equation}
    \mathcal{N} = \left[\partial J^{+}(c_{1}) \cup \partial J^{+}(c_{2})\right] \cap J^{-}(r_{1})  \cap J^{-}(r_{2}) . 
\end{equation}
$\mathcal{N}$ is bounded to the past by the ridge $\mathcal{R}_{pts}$ and to the future by the edge $\eth s_{pts}$ as depicted in the left image of Figure \ref{fig: spts and sreg}. Therefore, using Stokes' theorem and the fact that $\mathcal{N}$ is composed of the future directed causal horizons of $c_{1}$ and $c_{2}$, we have:
\begin{align}
    \area{\eth s_{pts}} - 2 \area{\mathcal{R}_{pts}} = \int_{\mathcal{N}} \theta \leq 0 .
\end{align}
Thus,
\begin{equation}
    S_{gen}(s''_{pts}) \leq \frac{1}{2 G_{N}} \area{\mathcal{R}_{pts}}.
\end{equation}

Recall that in the proof of the CWT we related the ridge of the regions-based scattering region to the mutual information $I(\hatit{V}_{1}; \hatit{V}_{2})$.
We can use another focusing argument to relate the ridges of the points-based scattering region and the regions-based one. 
\begin{figure}
    \centering
    \includegraphics[width=0.7\linewidth]{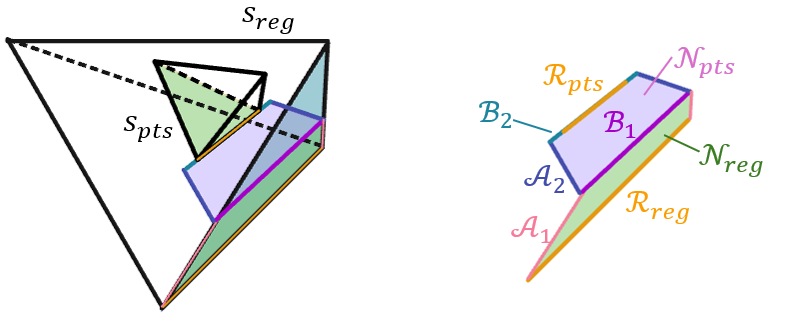}
    \caption{The left image shows the points based scattering region $s_{pts}$ inside $s_{reg}$ and the null surfaces defined for the proof of Theorem \ref{thm: Rpts < Rreg}. In the right image, we have extracted for clarity the relevant null surfaces and labeled their boundaries.}
    \label{fig:reg to pts}
\end{figure}
\begin{thm}
\label{thm: Rpts < Rreg}
    $\area{\mathcal{R}_{pts}} \leq \area{\mathcal{R}_{reg}}$ 
\end{thm}
\noindent Let $\mathcal{N}_{reg}$ be the null surface:
\begin{equation}
    \mathcal{N}_{reg} = \partial J^{+}(\ew{\hatit{V}_{1}})  \cap (J^{+}(c_{2}))^{c}  \cap  \text{cl}[s_{reg}]
\end{equation}
as depicted in Figure \ref{fig:reg to pts}. The additional intersection $ (J^{+}(c_{2}))^{c}$, means that the surface is cutoff to the future by the locus 
\begin{equation}
    \mathcal{B}_{1} = \partial J^{+}(v_{1}) \cap \partial J^{+}(c_{2}) \cap \text{cl}[ s_{reg}] . 
\end{equation}
Since $\mathcal{N}_{reg} \subset \partial J^{+}(\ew{\hatit{V}_{1}})$, we have:
\begin{equation}
    \area{\mathcal{B}_{1}} + \area{\mathcal{A}_{1}} - \area{\mathcal{R}_{reg}}  \leq 0
\end{equation}
where $\mathcal{A}_{1} := (\partial J^{-}(\ew{\hatit{W}_{1}})   \cup \partial J^{-}(\ew{\hatit{W}_{2}})) \cap \mathcal{N}_{reg}$ forms the remainder of the boundary of $\mathcal{N}_{reg}$.

Now define another null surface $\mathcal{N}_{pts}$ by
\begin{equation}
    \mathcal{N}_{pts} = \partial J^{+}(c_{2}) \cap  (J^{+}(c_{1}))^{c} \cap \text{cl}[s_{reg} ] .
\end{equation}
Note that $\mathcal{N}_{pts} \cap \partial J^{+}(v_{1}) = \mathcal{B}_{1}$. We also define another ridge $\mathcal{B}_{2}$ by:
\begin{equation}
    \mathcal{B}_{2} = \partial J^{+}(c_{2}) \cap \partial J^{+}(c_{1})  \cap J^{-}(\ew{\hatit{W}_{1}}) \cap J^{-}(\ew{\hatit{W}_{2}}) .
\end{equation}
Then since $\mathcal{N}_{pts} \subset \partial J^{+}(c_{2})$, we have decreasing area as we move to the future along $\mathcal{N}_{pts}$ i.e.
\begin{equation}
    \area{\mathcal{B}_{2}} + \area{\mathcal{A}_{2}} - \area{\mathcal{B}_{1}} \leq 0
\end{equation}
where $\mathcal{A}_{2} =  \left(\partial J^{-}(\ew{\hatit{W}_{1}}) \cap \partial J^{-}(\ew{\hatit{W}_{2}}) \right) \cap  \mathcal{N}_{pts}$ forms the remainder of the boundary of $\mathcal{N}_{pts}$. Combining the two results, we have:
\begin{equation}
    \area{\mathcal{B}_{2}} \leq \area{\mathcal{R}_{reg}} - \area{\mathcal{A}_{1} \cup \mathcal{A}_{2}} \leq  \area{\mathcal{R}_{reg}} .
\end{equation}
Now note that since $J^{-}(r_{1}) \cup J^{-}(r_{2}) \subset J^{-}(\ew{\hatit{W}_{1}}) \cap J^{-}(\ew{\hatit{W}_{2}}) $, $\mathcal{R}_{pts} \subset \mathcal{B}_{2}$ and hence
\begin{equation}
    \area{\mathcal{R}_{pts}} \leq \area{\mathcal{R}_{reg}} .
\end{equation}

We now have all the desired ingredients to establish our lower bound.
Combining Theorem \ref{thm: sgen(spts) leq ridge} and \ref{thm: Rpts < Rreg} with our result from the CWT \eqref{eq: CWT result} and \eqref{eq: S(emax) < S(a)}, we have:
\begin{equation}
    S_{gen}(e_{max}(s_{pts}'')) \leq S_{gen}(s_{pts}'')  \leq I(\hatit{V}_{1}; \hatit{V}_{2}).
\end{equation}
Meanwhile, by the same logic but requiring fewer intermediate steps we also have
\begin{equation}
    S_{gen}(e_{max}(s_{reg}'')) \leq S_{gen}(s_{reg}'')  \leq I(\hatit{V}_{1}; \hatit{V}_{2}).
\end{equation}
We have thus established the lower bounds in~\eqref{bound}.

\subsection{Upper Bounds from Entanglement Scattering Regions} \label{sec:upper}

We will now turn to the upper bound.  First, note that because $s_{ent}$ is the intersection of two entanglement wedges, it is a wedge as well. This step is simpler than in the $s_{pts/reg}$ case. A consequence of the CWT is that the scattering region $s_{reg}$ lies inside the connected entanglement wedge $\ew{\hatit{V}_{1} \cup \hatit{V}_{2}}$ \cite{May:2021nrl}. By the symmetry of our setup, we have $s_{reg} \subset \ew{\hatit{W}_{1} \cup \hatit{W}_{2}}$, so $s_{reg} \subset s_{ent}$ the entanglement scattering region~\eqref{s-ent} defined in~\cite{Leutheusser:2024yvf}.\footnote{The same is true for $s_{reg}''$ since this is the smallest wedge containing $s_{reg}$ and $s_{ent}$ is another such wedge containing $s_{reg}$.}
We can use a similar argument used to prove the CWT to relate the mutual information to the larger scattering region $s_{ent}$.

\begin{figure}
    \centering
    \includegraphics[width=0.5\linewidth]{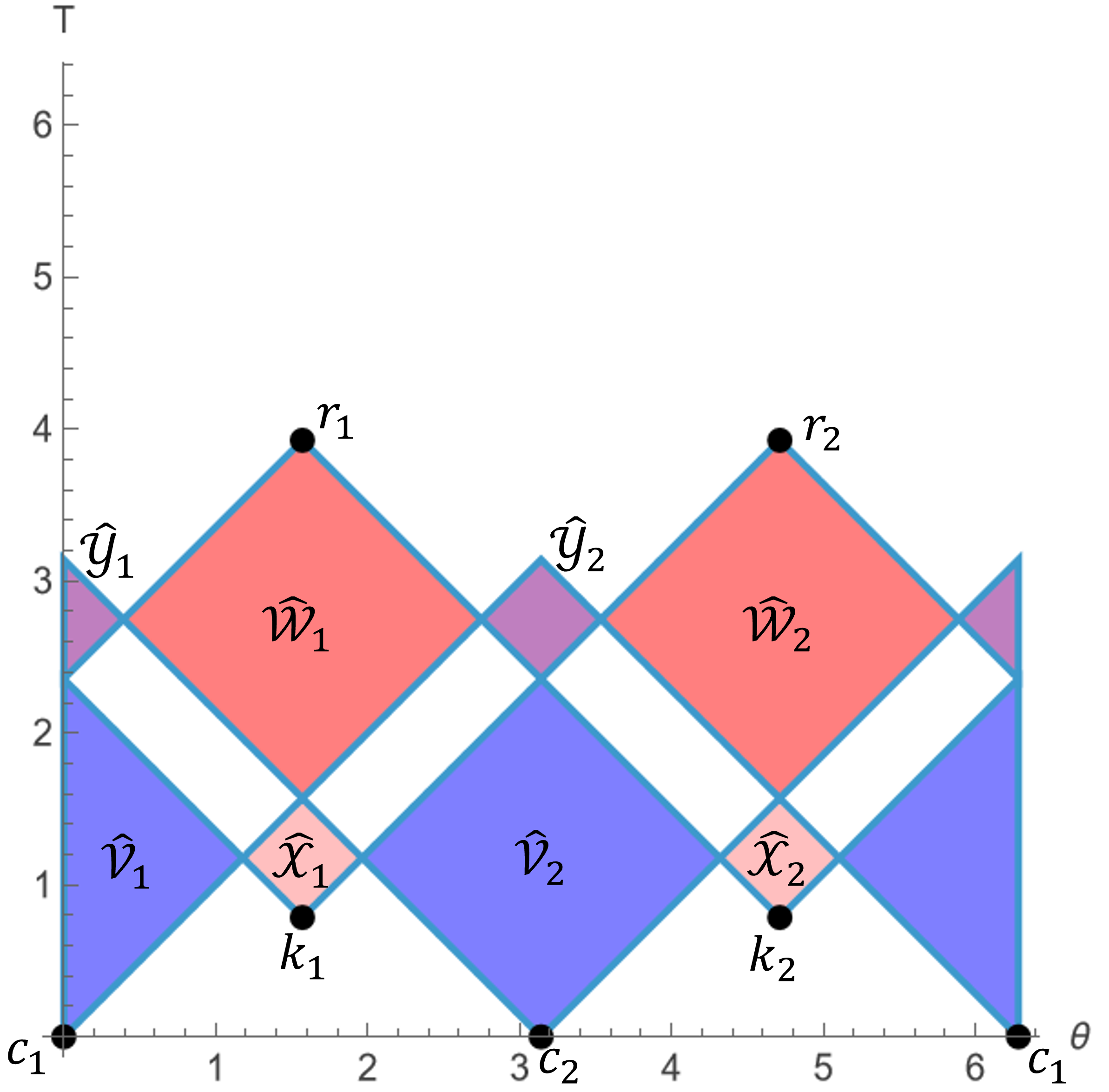}
    \caption{The decision regions $\hatit{V}$ and $\hatit{W}$ shown in blue and red respectively as well as the complement regions $\hatit{X}$ (pink) and $\hatit{Y}$ (purple). $k_{1}$ and $k_{2}$ are the antipodal points of $r_{2}$ and $r_{1}$ respectively. }
    \label{fig:complementX}
\end{figure}

For simplicity, we will assume that we are in a pure state and that $s_{pts}\neq\emptyset$ so that we are in the connected phase by the CWT. Then $\E{\hatit{V}_{1} \cup \hatit{V}_{2}}$ is composed of two disconnected extremal surfaces $\E{\hatit{X}_{1}}$ and $\E{\hatit{X}_{2}}$. In fact, $\E{\hatit{X}_{1}} \cup \E{\hatit{X}_{2}}  = \E{\hatit{X}}$ where  
\begin{equation}
    \hatit{X} = \hatit{X}_{1} \cup \hatit{X}_{2} = (\hatit{V}_{1} \cup \hatit{V}_{2})'
\end{equation}
is the spacelike complement of $\hatit{V}_{1} \cup \hatit{V}_{2}$ on the boundary as shown in Figure \ref{fig:complementX}.  Likewise, $\E{\hatit{W}_{1} \cup \hatit{W}_{2}}$ is also composed of two disconnected extremal surfaces, $\E{\hatit{Y}_{1}}$ and $\E{\hatit{Y}_{2}}$ where $\hatit{Y} = \hatit{Y}_{1} \cup \hatit{Y}_{2}$ is the spacelike complement of $\hatit{W}_{1} \cup \hatit{W}_{2}$ on the boundary. Using \eqref{eq: edge of intersection}, we have
\begin{align}\label{sentedge}
    \eth s_{ent} =  &\left(\bigcup_{i,j} \left(\partial J^{+}(\ew{\hatit{X}_{i}}) \cap \partial J^{-}(\ew{\hatit{Y}_{j}})  \cap s_{ent}\right)\right) \cup \nonumber \\
    &\left((\E{\hatit{X}_{1}} \cup \E{\hatit{X}_{2}}) \cap s_{ent}\right) \cup  \left((\E{\hatit{Y}_{1}} \cup \E{\hatit{Y}_{2}}) \cap s_{ent}\right) \\
    &\cup \left(\bigcup_{i,j} \left(\partial J^{-}(\ew{\hatit{X}_{i}}) \cap \partial J^{+}(\ew{\hatit{Y}_{j}})  \cap s_{ent}\right)\right) . \nonumber 
\end{align}
Depending on the topology of the intersection, the structure of edges of $s_{ent}$ can be more complicated than for $s_{pts}$ or $s_{reg}$. However, we still have similar control over the signs of expansions. 
In contrast to $s''_{pts}$ and $s''_{reg}$, the wedge $s_{ent}$ is normal because its future horizon is generated by the future lightsheets from $\E{\hatit{X}_{1}}$ and $\E{\hatit{X}_{2}}$ while its past horizon is generated by the past lightsheets from $\E{\hatit{Y}_{1}}$ and $\E{\hatit{Y}_{2}}$.

\begin{thm} $S_{gen}(s_{ent}) \geq I(\hatit{V}_{1}; \hatit{V}_{2})$.
\label{thm: S(sent) > I}
\end{thm}
\noindent We will prove this theorem in two steps. First, we will demonstrate the structure of the proof under simplifying assumptions. Then we will use focusing to relate the wedge entropy to the area of an intermediate ridge for the more general configuration. This ridge upper bounds the mutual information, and we include the proof in appendix~\ref{app:ridge}. The proof in the simplest example was demonstrated in~\cite{Lima:2025dtj}  in the context of studying~\cite{Leutheusser:2024yvf} and trying to build a generalized connected wedge theorem (GCWT) with converse.

\begin{figure}
    \centering
    \includegraphics[width=0.55\linewidth]{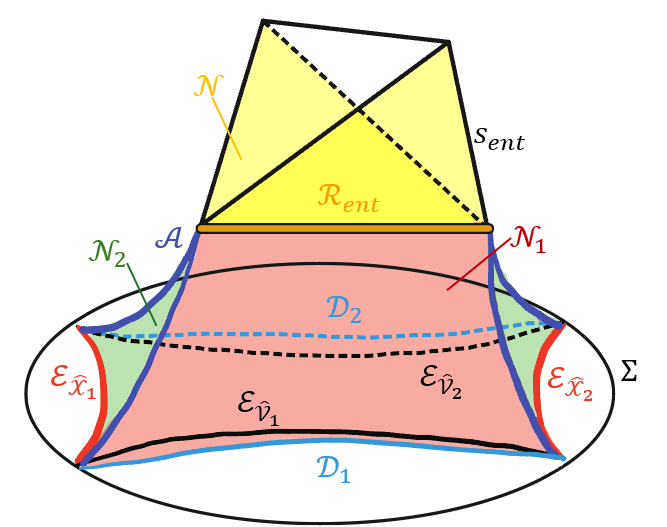}
    \caption{The null surfaces $\mathcal{N}$ (yellow), $\mathcal{N}_{1}$ (dark red) and $\mathcal{N}_{2}$ (green) and their boundaries defined in the proof of Theorem \ref{thm: S(sent) > I}.}
    \label{fig:sentpf}
\end{figure}

Let $\Sigma$ be a Cauchy slice such that $\E{\hatit{X}_{1} \cup \hatit{X}_{2}} \subset \Sigma$ satisfying the maximin condition. Let us first consider the case where $s_{ent}$ lies entirely above $\Sigma$. Then
$(\E{\hatit{X}_{1}} \cup \E{\hatit{X}_{2}}) \cap s_{ent} = \emptyset$, and by assumption the ridge 
\begin{equation}
    \mathcal{R}_{ent} =  \left[\partial J^{-}(\ew{\hatit{Y}_{1}}) \cap \partial J^{-}(\ew{\hatit{Y}_{2}})\right] \cap (J^{+}(\ew{\hatit{X}_{1}}))^{c} \cap (J^{+}(\ew{\hatit{X}_{2}}))^{c} 
    \label{eq: Re def}
\end{equation}
is also to the future of $\Sigma$. To simplify things even further for this warm-up we will let  $(\E{\hatit{Y}_{1}} \cup \E{\hatit{Y}_{2}}) \cap s_{ent} = \emptyset$. 
Then our entanglement scattering region will look like a tetrahedron as drawn in~\eqref{fig:sentpf} and $\eth s_{ent}$ will only consist of the first line in~\eqref{sentedge}.
In this case, we can use a very similar argument as in Theorem \ref{thm: sgen(spts) leq ridge}  to show that $ S_{gen}(s_{ent}) \geq \area{\mathcal{R}_{ent}}$. We define:
\begin{equation}
     \mathcal{N} = \left[\partial J^{-}(\ew{\hatit{Y}_{1}}) \cup \partial J^{-}(\ew{\hatit{Y}_{2}})\right] \cap \text{cl}[s_{ent}] .
\end{equation}
Then since $\mathcal{N}$ is generated by past directed lightsheets, Stokes' Theorem gives:
\begin{equation}
    2 \area{\mathcal{R}_{ent}} - \area{\eth s_{ent}} = \int_{\mathcal{N}} \theta \leq 0
\end{equation}
i.e.
\begin{equation}
\label{eq: Sgen(sent) > Re}
    S_{gen}(s_{ent}) \geq \frac{1}{2 G_{N}} \area{\mathcal{R}_{ent}}.
\end{equation}

The next step is to relate the ridge $\mathcal{R}_{ent}$ to the mutual information. Define the null surface:
\begin{equation}\label{N1}
    \mathcal{N}_{1} = \partial \left[ J^{-}(\ew{\hatit{Y}_{1}}) \cap  J^{-}(\ew{\hatit{Y}_{2}})\right] \cap (J^{+}(\ew{\hatit{X}_{1}}))^{c} \cap (J^{+}(\ew{\hatit{X}_{2}}))^{c} \cap J^{+}(\Sigma) .
\end{equation}
$\mathcal{N}_{1}$ meets $\Sigma$ at $\mathcal{D} =\mathcal{D}_{1} \cup \mathcal{D}_{2}$ where:
\begin{equation}
    \mathcal{D}_{i} = \partial J^{-}(\ew{\hatit{Y}_{i}})  \cap (J^{+}(\ew{\hatit{X}_{1}}))^{c} \cap (J^{+}(\ew{\hatit{X}_{2}}))^{c} \cap \Sigma .
\end{equation}
Note that $\mathcal{D}_{i}$ is homologous to $\hatit{V}_{i}$ so 
\begin{equation}
    \area{\mathcal{D}} \geq \area{\E{\hatit{V}_{1}} \cup \E{\hatit{V}_{2}}} .
    \label{eq: mutual info step}
\end{equation}
In addition, we can define a second null surface $\mathcal{N}_{2}$ by:
\begin{equation}\label{N2}
    \mathcal{N}_{2} = \partial \left[J^{+}(\ew{\hatit{X}_{1}}) \cup  J^{+}(\ew{\hatit{X}_{2}})\right] \cap J^-(\ew{\hatit{Y}_{1}})\cap J^-(\ew{\hatit{Y}_{2}}).
    \end{equation}

As in the proof of the CWT, $\mathcal{N}_{1}$ and $\mathcal{N}_{2}$ are joined by seams $\mathcal{A}$. $\mathcal{N}_{1}$ and $\mathcal{N}_{2}$ are analogous to the lift and slope, except that here the ``lift'' $\mathcal{N}_{1}$ is generated by a past-directed lightsheet and the ``slope'' $\mathcal{N}_{2}$ is generated by a future-directed null sheet. 
Hence, the inequality is reversed compared to the previous case:
\begin{align}
    \area{\mathcal{D}} - \area{\mathcal{A}} - 2 \area{\mathcal{R}_{ent}} &= \int_{\mathcal{N}_{1}} \theta \leq 0 \\
    \area{\mathcal{A}} - \area{\E{\hatit{X}_{1} \cup \hatit{X}_{2}}} &= \int_{\mathcal{N}_{2}} \theta \leq 0 .
\end{align}
Adding the two equations, and rearranging we have:
\begin{equation}
    \area{\mathcal{D}}  - \area{\E{\hatit{X}_{1} \cup \hatit{X}_{2}}} \leq 2 \area{\mathcal{R}_{ent}} . 
\end{equation}
Then \eqref{eq: mutual info step} implies that:
\begin{equation}
    I(\hatit{V}_{1}; \hatit{V}_{2}) \leq \frac{1}{2 G_{N}}\area{\mathcal{R}_{ent}} .
    \label{eq: I < Re}
\end{equation}
Putting \eqref{eq: Sgen(sent) > Re} and \eqref{eq: I < Re} together, we have the desired result:
\begin{equation}
    I(\hatit{V}_{1}; \hatit{V}_{2}) \leq S_{gen}(s_{ent})
\end{equation}
albeit in a rather restricted setting. 

\begin{figure}
    \centering
    \includegraphics[width=0.9\linewidth,trim={1em 1.5em 1em 1em}]{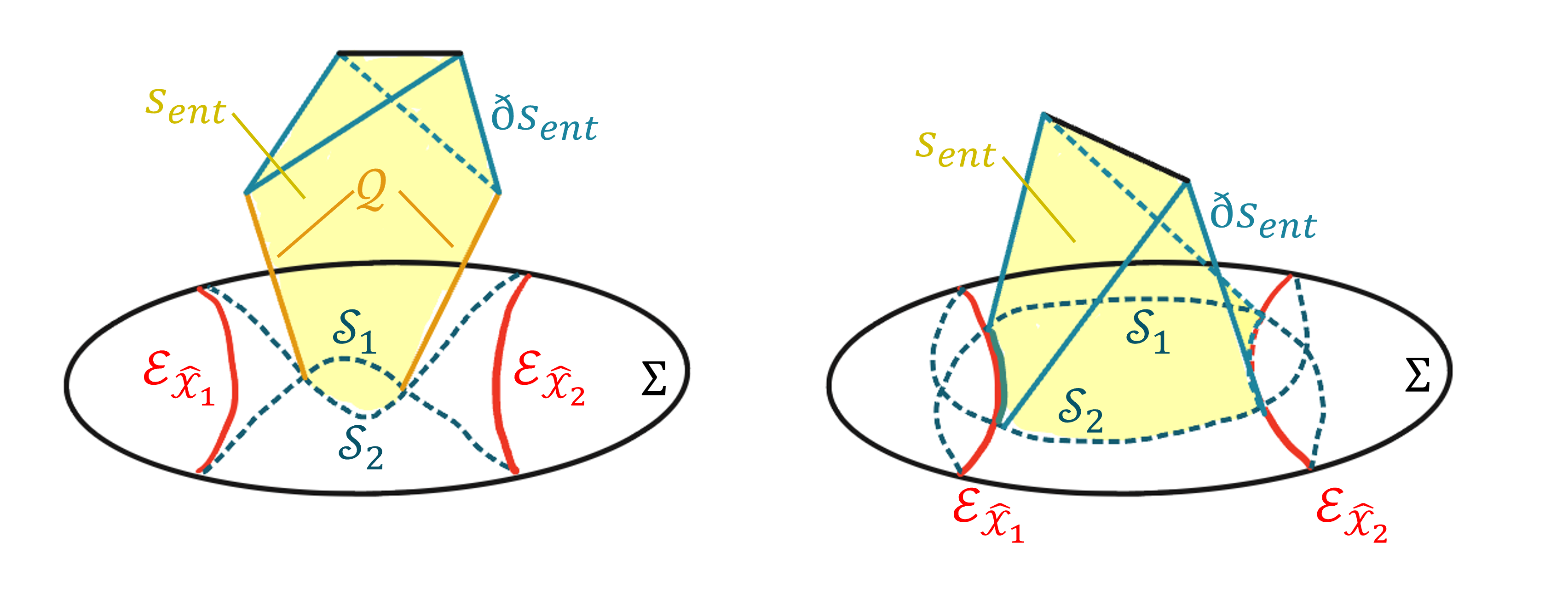}
    \caption{\textit{Left}:  A configuration with $\E{\hatit{X}_{1}} \cap s_{ent} = \emptyset$ and $\E{\hatit{X}_{2}} \cap s_{ent} = \emptyset$. \textit{Right}: A configuration with $\E{\hatit{X}_{1}} \cap s_{ent} \neq \emptyset$ and $\E{\hatit{X}_{2}} \cap s_{ent} \neq \emptyset$.   The part of $s_{ent}$ which lies above the Cauchy slice is shaded in yellow. The edge is indicated with a blue outline. The part of the ridge $\mathcal{Q}$ which lies above the Cauchy slice is indicated in orange. }
    \label{fig: cases}
\end{figure}

Let us now turn to the more general case where the ridge~\eqref{eq: Re def} can dip below $\Sigma$. Some examples of what the scattering region can look like are included in figure~\ref{fig: cases}, though it can even be more complicated if the ${\cal E}_{\hat\Y}$ also dip below $\Sigma$, which does not seem straightforward to rule out. Rather than demonstrate the proof in a case-by-case fashion, we will begin by co-opting a result from~\cite{Lima:2025dtj}: that $s_{ent}\neq \emptyset$ whenever the wedges are connected and that this region cuts across $\ew{\V_1\cup \V_2}$. In appendix~\ref{app:ridge} we will show that we can prove an upper bound in terms of a modified composite ``ridge''
\be\label{ridgeupper}
 I(\hatit{V}_{1}; \hatit{V}_{2}) \leq \sum_{i=1}^2 \frac{1}{4 G_{N}}\area{\mathcal{R}_{ent,i}}.
\ee
We will explain the definition of the ridge, including why it has an additional label, shortly. 
In order to prove Theorem~\ref{thm: S(sent) > I} we will take~\eqref{ridgeupper} as a given and instead prove the relation
\be\label{sentupper}
\sum_{i=1}^2 \frac{1}{4 G_{N}}\area{\mathcal{R}_{ent,i}} \le S_{gen}(s_{ent}).
\ee
 For now we just note that the proof of~\eqref{ridgeupper} proceeds by focusing the ``ridge'' 
 towards $\Sigma$. Meanwhile the proof of~\eqref{sentupper} proceeds by focusing the edge of the wedge towards the ridge, as we will now show.

Let us now define the ridge surfaces $\mathcal{R}_{ent,i}$. It will be useful to split these into a few parts. First we define
\begin{equation}
    {\cal R}_{+} =  \left[\partial J^{-}(\ew{\hatit{Y}_{1}}) \cap \partial J^{-}(\ew{\hatit{Y}_{2}})\right] \cap (J^{+}(\ew{\hatit{X}_{1}}))^{c} \cap (J^{+}(\ew{\hatit{X}_{2}}))^{c} \cap J^+(\Sigma)
    \label{eq: rp}
\end{equation}
which is nearly identical to~\eqref{eq: Re def}, up to the restriction that we are in the future of $\Sigma$, which was automatic in our discussion above. Similarly, we define
\begin{equation}
    {\cal R}_{-} =  \left[\partial J^{+}(\ew{\hatit{Y}_{1}}) \cap \partial J^{+}(\ew{\hatit{Y}_{2}})\right] \cap (J^{-}(\ew{\hatit{X}_{1}}))^{c} \cap (J^{-}(\ew{\hatit{X}_{2}}))^{c} \cap J^-(\Sigma).
    \label{eq: rm}
\end{equation}
Finally let
\begin{align}\label{eq:si}
    \mathcal{S}_{i} &= [\partial J^{-}(\ew{\hatit{Y}_{i}}) \cup \partial J^{+}(\ew{\hatit{Y}_{i}}) ]\cap s_{ent} \cap \Sigma .
\end{align}
Our composite ``ridge'' will then be made up of the pieces  
\begin{equation}
    \mathcal{R}_{ent,i} =  {\cal R}_+\cup {\cal R}_-\cup  \mathcal{S}_{i}.
    \label{eq: Rei}
\end{equation}
We can see these extra components are designed to handle the case where the ${\cal E}_{\Y_i}$ dip below $\Sigma$, otherwise we would only need configurations like the type illustrated in figure~\ref{fig: cases}, where $\eth s_{ent}$ is strictly to the future of $\Sigma$ even if $s_{ent}$ isn't. 

By construction, $s_{ent}$ is to the future of ${\cal R}_+$ and to the past of ${\cal R}_-$, while the ${\cal S}_i$ make up the boundary of the intersection of $s_{ent}$ with $\Sigma$. We now want to focus the edge~\eqref{sentedge} towards this ridge to establish~\eqref{sentupper}. The key ingredient is analogous to what we used in section~\ref{sec:lower}. Namely, not only do we have control over the signs of the expansions at the edges of the wedge, we also know what the sign is along the entire Cauchy horizon. We can thus split $\eth s_{ent}$ into segments to the future of $\Sigma$:
\begin{align}\label{sentedge+}
    \eth s_{ent} \cap J^+(\Sigma) =  &\left(\bigcup_{i,j} \left(\partial J^{+}(\ew{\hatit{X}_{i}}) \cap \partial J^{-}(\ew{\hatit{Y}_{j}})  \cap s_{ent}\right)\right) \nonumber \\ &\cup  [\left((\E{\hatit{Y}_{1}} \cup \E{\hatit{Y}_{2}}) \cap s_{ent}\right) \cap J^+(\Sigma)] 
\end{align}
and to the past of $\Sigma$
\begin{align}\label{sentedge-}
    \eth s_{ent} \cap J^-(\Sigma) =  &
    \left(\bigcup_{i,j} \left(\partial J^{-}(\ew{\hatit{X}_{i}}) \cap \partial J^{+}(\ew{\hatit{Y}_{j}})  \cap s_{ent}\right)\right) \nonumber \\ & \cup [ \left((\E{\hatit{Y}_{1}} \cup \E{\hatit{Y}_{2}}) \cap s_{ent}\right) \cap J^-(\Sigma)] .
\end{align}
The part of the edge on $\Sigma$ will not contribute to the ridge. Now note that for both of the terms in~\eqref{sentedge+} the generators of $H^-(s_{ent})$ coming out of the edge will be generators of $\p J^-(y_i)$. Similarly, for both sets of terms in~\eqref{sentedge-}, the generators of $H^+(s_{ent})$ emerging from these edge components will be generators of $\p J^+(y_i)$. The former will thus focus back towards ${\cal R}_+$ and some segment of the ${\cal S}_i$, while the latter will focus forward to ${\cal R}_-$ and the remainder of the ${\cal S}_i$ so that from our area theorem we have 
\be
\area{\eth s_{ent} \cap J^+(\Sigma)}+\area{\eth s_{ent} \cap J^-(\Sigma)} \ge \sum_{i=1}^2 \area{{\cal R}_{ent,i}}.
\ee
The remaining component of $\eth s_{ent}$ contained in $\Sigma$ only increases the area, and we have thus established~\eqref{sentupper}. Assuming~\eqref{ridgeupper}, we have thus established Theorem~\ref{thm: S(sent) > I}.

\begin{figure}
    \centering
    \includegraphics[width=0.75\linewidth]{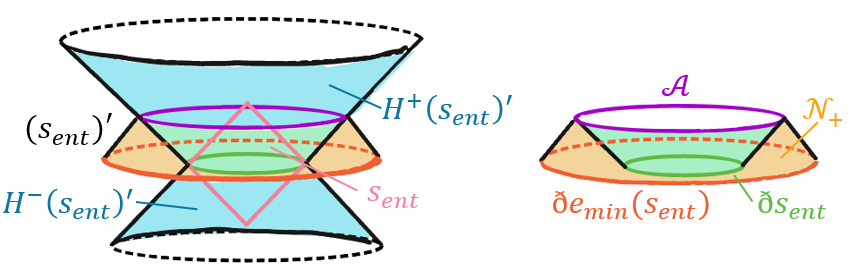}
    \caption{\textit{Left:} The wedge $s_{ent}$ outlined in pink (note that to avoid clutter, we have not drawn the tetrahedral shape of $s_{ent}$) and the Cauchy horizons $H^{\pm}(s_{ent}')$ in blue. The edge of $e_{min}(s_{ent})$ is a spacelike curve that lies outside the edge of $s_{ent}$. \textit{Right:} The null surface $\mathcal{N}_{+}$ defined in the proof of Theorem \ref{thm: S(sent) leq S(emin)} and its boundaries.}
    \label{fig:eminpf null surf}
\end{figure}

 The last step in establishing the upper bound~\eqref{bound} is to relate these quantities to $S_{gen}(e_{min}(s_{ent}))$. Since the wedge $s_{ent}$ is normal, instead of antinormal like $s_{pts}$ and $s_{reg}$, we are able to use the property that $e_{min}(a)$ is marginal or extremal on $\eth e_{min}(a) - \eth a$ to relate the entropy of $s_{ent}$ to the entropy of its min entanglement wedge $e_{min}(s_{ent})$.
\begin{thm}
\label{thm: S(sent) leq S(emin)}
   $S_{gen}(s_{ent}) \leq S_{gen}(e_{min}(s_{ent}))$ .
\end{thm}
\noindent Because $e_{min}(s_{ent})$ is larger than $s_{ent}$, the edge of $e_{min}(s_{ent})$ is generally outside of the edge of $s_{ent}$ and we know $\eth e_{min}(s_{ent}) - \eth a$ is contained in $\text{cl}(a')$. Let us define a null surface $\mathcal{N}_{+}$ by firing future-directed null geodesics orthogonally from $\eth e_{min}(s_{ent})$ to $H^{+}(s_{ent}')$ as shown in Figure \ref{fig:eminpf null surf}. Let $\mathcal{A}$ be the intersection of this null surface with the future Cauchy horizon of $a'$: $\mathcal{A} = \mathcal{N}_{+} \cap H^{+}(s_{ent}')$. By property 4 in Theorem \ref{thm: properties of emin and emax}, $\eth e_{min}(s_{ent})$ is extremal at all points $p \in \eth e_{min} \cap a'$. Therefore, $\mathcal{N}_{+}$, as an orthogonal null sheet to an extremal surface, has $\theta \leq 0$. Consequently, 
\begin{equation}
    \area{\mathcal{A}} \leq \area{\eth e_{min} (s_{ent})} .
\end{equation}
Now consider the null surface formed by the Cauchy horizon $H^{+}(s_{ent}')$ between $\mathcal{A}$ and $\eth s_{ent}$. Note that $H^{+}(a')$ is generated by the past-moving lightsheets from the extremal surfaces $\E{\hatit{Y}_{1}}$ and $\E{\hatit{Y}_{2}}$. Consequently, $\theta \leq 0$ as we move to the past from $\mathcal{A}$ to $\eth s_{ent}$. Hence, 
\begin{equation}
    \area{\eth s_{ent}} \leq \area{\mathcal{A}}.
\end{equation}
We thus have
\begin{equation}
    S_{gen}(s_{ent}) \leq S_{gen}(e_{min}(s_{ent})).
\end{equation}
Putting together Theorem \ref{thm: S(sent) > I} and \ref{thm: S(sent) leq S(emin)} then gives us
\begin{equation}
    I(\hatit{V}_{1}; \hatit{V}_{2})  \leq S_{gen}(s_{ent}) \leq S_{gen}(e_{min}(s_{ent})) ,
\end{equation}
 the desired upper bound~\eqref{bound}.

\section{A Bulk Generalization of the CWT}\label{sec:bulkcwt}
We will now use the constructions of the previous section to establish a generalization of the connected wedge theorem that moves the decision regions into the bulk in a manner that will be amenable to taking a flat limit in section~\ref{sec:flat}. 
The goal is to identify a suitable set of bulk decision regions $v_{i}$ for which the focusing methods of the previous sections will still work to show a generalized CWT:
\be\label{sptsconn3}
s_{pts}\neq\emptyset~\Rightarrow~ e_{min}(v_1\Cup v_2)~{\rm connected}.
\ee
We will prove this statement by showing that no disconnected candidate can be the $e_{min}(v_1\Cup v_2)$, rather than  comparing to a particular alternate candidate for $e_{min}$.  As such, we won't be able to supply a lower bound on the mutual information the same way we were able to in the previous section. However, we can still apply these steps to the original CWT problem.

\paragraph{Warmup: Reproducing the CWT}
As a primer to the rest of this section, let us rederive the original connected wedge theorem using the steps we will take for more general choices of $v_i$ in the following subsections. 
First, recall from the discussion below~\eqref{eq:ecw} that for compressible states (or in thee semi-classical limit) in boundary regions in asymptotically AdS, \be
\label{eq:e(cw)}
\ew{\hatit{V}_i}=e_{min}(c_W(\hatit{V}_i))=e_{max}(c_W(\hatit{V}_i)). 
\ee 
Consequently, the Connected Wedge Theorem can be rewritten as:
\begin{equation}\label{eq:conncw}
    s_{pts}\neq\emptyset~\Rightarrow~ e_{min}(\cw{\hatit{V}_{1} \cup \hatit{V}_{2}})~{\rm connected}.
\end{equation}
Now since the $\hatit{V}_{i}$ are spacelike separated $\cw{\hatit{V}_{1} \cup \hatit{V}_{2}} = \cw{\hatit{V}_{1}} \Cup \cw{\hatit{V}_{2}}$, we obtain \eqref{sptsconn3} where we let $v_{i} = \cw{\hatit{V}_{i}}$.
In particular, the proof of \eqref{sptsconn3} in this case should combine the proof of the Connected Wedge Theorem with the proof of~\eqref{eq:e(cw)}. 

Let $a=\cw{\V_1}\Cup \cw{\V_2}$. This is just the normal union since the regions are spacelike separated. Assume for contradiction that its entanglement wedge $e_{min}(a)$ is given by two disconnected components $g_{i}$ with $v_{i} \subset g_{i}$. In particular, this requires that $g_{1}$ and $g_{2}$ be spacelike so $\area{\eth g} =\area{g_{1}} + \area{g_{2}}$.
According to the third condition of Definition \ref{def:min}, $g' \cap \tilde{a}'$ admits a Cauchy slice $\Sigma'$ such that there does not exist a wedge $h \neq g$ such that:
\begin{enumerate}
    \item $g \subset h \subset \tilde{a}'$
    \item $\eth h \subset \Sigma'$
    \item $\area{\eth h} \leq \area{\eth g}$
    \item $\eth h - \eth g$ is compact
\end{enumerate}
We will show a contradiction by finding such an $h$.  Note that $g \subset \tilde{a}'$ implies that $\eth g \subset \Sigma'$ and $\eth \tilde{a} \subset \Sigma'$. Let $\Sigma$ be a Cauchy slice which contains $\Sigma'$. Since $\eth a$ is spacelike to $\eth g$, we may choose $\Sigma$ such that it contains $\eth a$.
Theorem~\ref{minedge} along with the property that the expansion is non positive on a causal horizon, implies that $\theta\le 0$ at the edge $\eth g$.  Therefore, we can follow the same focusing steps as in section~\ref{sec:cwt}.  Meanwhile, the past light sheets from $\p J^-(r_i)$ will necessarily intersect $\eth g$ since $\cw{\V_1}\subset g_i$. Furthermore, this nesting implies that $\p J^+(g_1)\cap \p J^+(g_1)$ will be strictly to the past of $\p J^+(\cw{\V_1})\cap \p J^+(\cw{\V_2})$ and thus $s_{pts}\neq\emptyset$ implies that the ridge 
\begin{equation}
\label{eq: ridge def2}
    \mathcal{R}_{g} = \partial J^{+}(g_1) \cap \partial J^{+}(g_2) \cap J^{-}(r_1)  \cap J^{-}(r_2) 
\end{equation}
is non empty. As such we can follow analogous steps as in the original CWT proof~\cite{May:2019odp} to show that the curves 
\be
{\cal C}_i=\partial J^{-}(r_i) \cap J^{-}[\partial J^{+}(g_{1} \cup g_{2})] \cap \Sigma
\ee
and
\begin{equation}
    \mathcal{D}_{i} = \eth g_{i}  \cap J^{-}(r_1)  \cap J^{-}(r_2) 
\end{equation}
obey
\be
2\area{\mathcal{R}_{g}}\le \sum_{i=1}^2 [\area{\mathcal{D}_{i}}-\area{{\cal C}_i}].
\ee
This would be true for any Cauchy slice containing the $\eth g_i$. 
If we let 
\begin{equation}
    \eth h = \bigcup_{i=1,2} \mathcal{C}_{i} \cup [\eth g \cap (J^{-}(r_{1}))^{c} \cap (J^{-}(r_{2}))^{c}]
\end{equation} we have identified a larger wedge $g\subset h$ with $\area{\eth h} \leq \area{\eth g}$, hence violating condition 3 of Definition \ref{def:min}.

\begin{figure}
    \centering
    \includegraphics[width=0.9\linewidth]{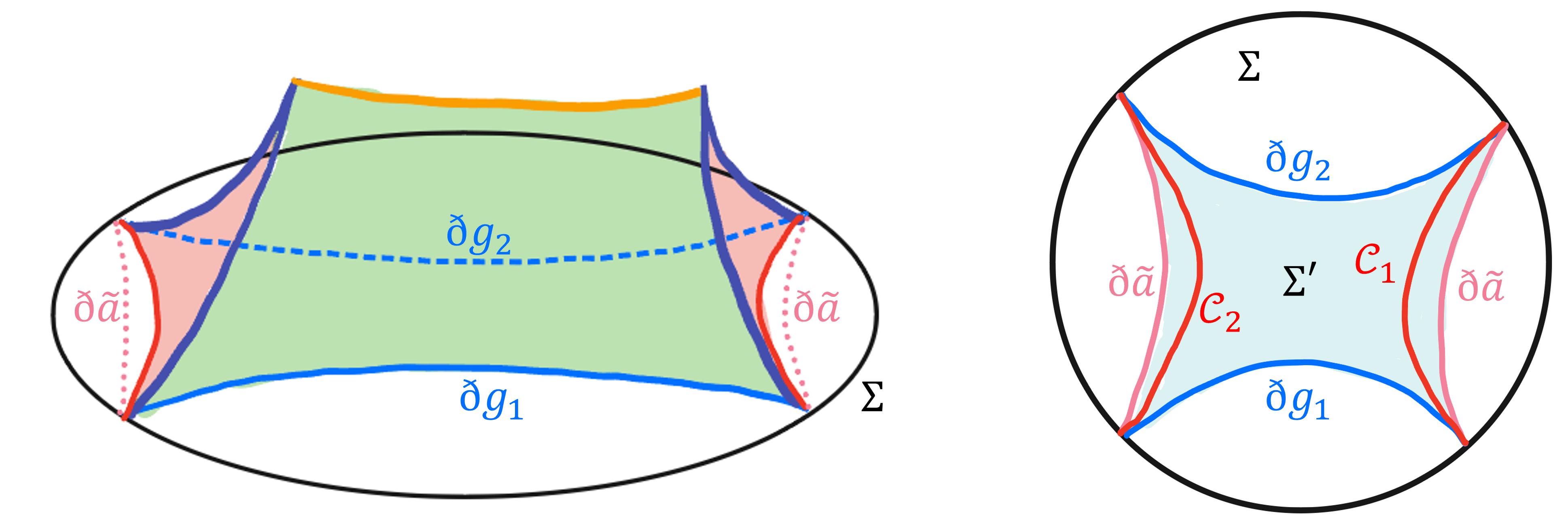}
    \caption{\textit{Left:}  Lift (green) and slope (red) used to find contradiction surface $\mathcal{C}$. In this case, we have $\mathcal{C} = \eth h$. \textit{Right:} View of Cauchy slice $\Sigma'$, including the edge of the fundamental complement $\eth a$. In this example, we find $J^{-}(r_{i})$ hit $\Sigma$ within $\Sigma'$ i.e. outside the fundamental complement.} 
    \label{fig:warmup}
\end{figure}

Now to finish the proof we just need to show that $h \subset \tilde{a}'$ i.e. $\eth h \subset \Sigma'$. Recall that the fundamental complement $\tilde a$ of a wedge $a$ is the causal wedge of its spacelike complement. The causal wedge will contain all the timelike curves with infinite proper length. We see from figure~\ref{fig:complementX} that $a'$ intersects the conformal boundary at the $\hat \X_i$ so $\tilde a=\cw{\hatit{X}_1}\Cup \cw{\hatit{X}_2}$. Additionally, from the boundary causal structure shown in figure~\ref{fig:complementX}, we see that $a \subset g \subset \tilde{a}'$ implies that $\eth a$ and $\eth g$ must meet at the boundary as depicted on the right side of Figure \ref{fig:warmup}. If we let $k_{i}$ be the past boundary point of $\hatit{X}_{i}$, then $\eth \tilde{a} = \partial J^{+}(k_{1}) \cup \partial J^{+}(k_{2}) \cap \Sigma$. Note that $k_{1}$ is antipodal to $r_{2}$ (i.e. future directed lightrays from $k_i$ on the boundary will converge at $r_i$, at least for any state that doesn't modify the boundary causal structure). Consequently, $\p J^+(k_i)$ in the bulk will be tangent or to the future of $\p J^-(r_i)$, and hence $\partial J^{-}(r_{2}) \cap \Sigma \in \Sigma$. Following the same logic for $k_{2}$ and $r_{1}$, we have found $\eth h \subset \Sigma'$. We thus have found the wedge $h$ which contradicts condition 3 of Definition \ref{def:min}.

Note that we could have followed the exact same proof as above setting $v_{i} = \ew{\hatit{V}_{i}}$ instead. In this case, the requirement that $\theta \leq 0$ at $\eth g$ follows from Theorem \ref{minedge} and the fact that $\eth\ew{\hatit{V}_{i}}$ is an extremal surface. 

\vspace{1em}\noindent
In the next two subsections we will consider two different families of decision regions $v_i$, in turn. These both have the feature that they can be restricted to the above example in certain limits. However, for the examples that follow we will need a small modification. Namely, when finding the wedge $h$ for contradiction, we will seemingly need to relax property 1 that $g \subset h \subset \tilde{a}'$ for our bulk generalizations. Since the definitions of the max- and min- entanglement wedges are still somewhat in flux~\cite{Bousso:2022hlz,Bousso:2023sya,Bousso:2024iry,Bousso:2025fgg}, we will content ourselves for the remainder of this section to make the weaker claim using the original definition in~\cite{Bousso:2023sya} which we will label  $e_{min,0}$:
\begin{defn}[Min$_0$-Entanglement Wedge]\label{def:min0}
    Given a wedge $a$, the min$_0$-entanglement wedge $e_{min,0}(a)$ is the intersection:
    \begin{equation}\label{eq:intmin}
        e_{min,0}(a) = \cap_{g\in G(a)} g
    \end{equation}
    of all wedges $g \in G(a)$, where $G(a)$ is the set of wedges satisfying:
    \begin{enumerate}
        \item $a \subset g$ and $\tilde\eth g=\tilde \eth a$
        \item $g$ is normal
        \item $g'$ admits a Cauchy slice $\Sigma'$ such that for any wedge $h \neq g$ such that $g \subset h$, $\eth h \subset \Sigma'$, and $\eth h - \eth g$ is compact we have:
        \begin{equation}
            S_{gen}(g) < S_{gen}(h) .
            \label{eq: sgen g < sgen h2}
        \end{equation}
    \end{enumerate}
\end{defn}
\noindent Similar modifications appear for the max$_0$-entanglement wedges. Note that in particular there is no longer a constraint on the wedge $h$ being spacelike to the fundamental complement.

With that in mind, let us now proceed to two classes of bulk generalizations.

\subsection{Entangled Pairs of Causal Diamonds}\label{sec:pairs} 

For our first example of a bulk generalization, we will take our $v_i$ to be
\be\label{vid}
v_i=(J^+(c_i)\cap J^-(b_i))'',~~~b_i\in \p J^-(r_1)\cap \p J^-(r_2)  \cap J^{+}(c_{i})
\ee
namely wedges constructed from causal diamonds with past end points being the same $c_i$, and future end points on the curve ${\cal R}=\p J^-(r_1)\cap \p J^-(r_2) $  defined by the intersections of the past lightcones of $r_1$ and $r_2$. This gives us a continuous family of causal diamonds which includes the original $\cw{\V_i}$ in the limit we send $b_i\rightarrow \p \tilde{M}$ towards the conformal boundary. 

\begin{figure}
    \centering
    \includegraphics[width=0.9\linewidth]{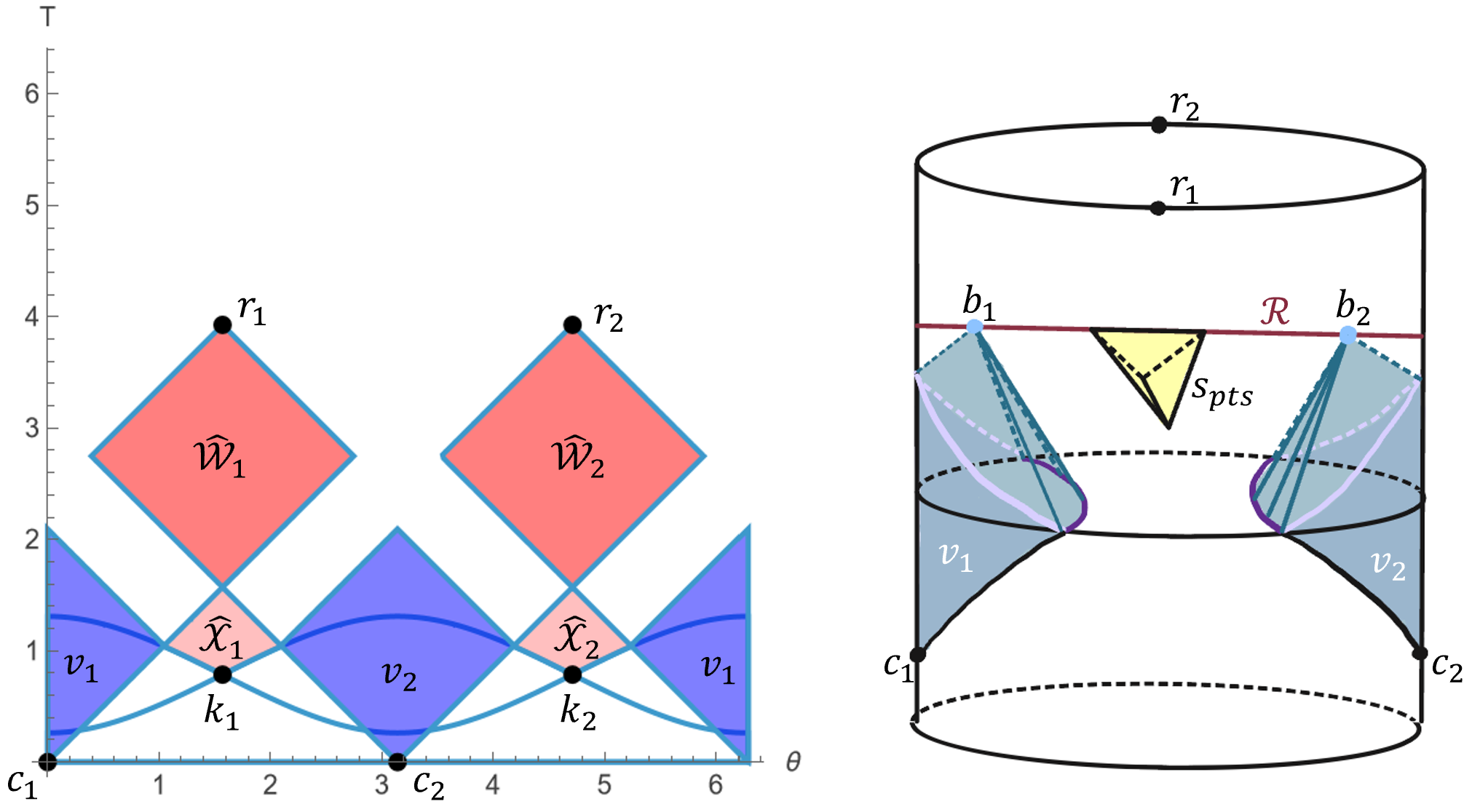}
    \caption{\textit{Left:} Boundary causal structure. The blue curves indicate the intersection of $\partial J^{-}(b_{1})$ and $\partial J^{-}(b_{2})$ with the boundary. Hence, the pink region gives the intersection of the fundamental complement with the boundary. The $v_{i}$ intersect the boundary at a diamond contained inside the original $\hatit{V}_{i}$. \textit{Right: } Bulk depiction of causal diamonds $v_{i}$ and the scattering region $s_{pts}$} 
    \label{fig:direg}
\end{figure}

Again letting $a = v_{1} \Cup v_{2}$ and assuming for contradiction that $e_{min}(a) := g$ is composed of two disconnected components $g_{i}$ with $v_{i} \in g_{i}$, we can follow almost the exact same steps as above to construct a wedge $h$ containing $g$ with $\area{\eth h} \leq \area{\eth g}$. 
As before, for any point $p \in \eth g - \eth a$, Theorem \ref{minedge} implies that $\eth g$ is extremal at $p$. Meanwhile, for any $p \in \eth a$, $p$ lies on a causal horizon $\partial J^{+}(c_{i})$. Consequently, $\theta \leq 0$ at $\eth g$ as needed for our focusing steps. Additionally, since $\eth g_{i}$ is outside of $\eth v_{i}$, the inward directed null sheets $\partial J^{+}(g_{i})$ meet to the past of the intersection between the $\partial J^{+}(c_{i})$, and hence the ridge $\mathcal{R}_{g}$ is non empty.  Thus, so long as the $v_i$ are spacelike separated\footnote{In the previous proof we knew that $\ew{v_i}$ were spacelike separated. We will return to fill this gap at the end of this section.} the focusing argument to find $h$ will follow nearly the same as in the original case considered above, with one modification.

\begin{figure}
    \centering
    \includegraphics[width=0.9\linewidth]{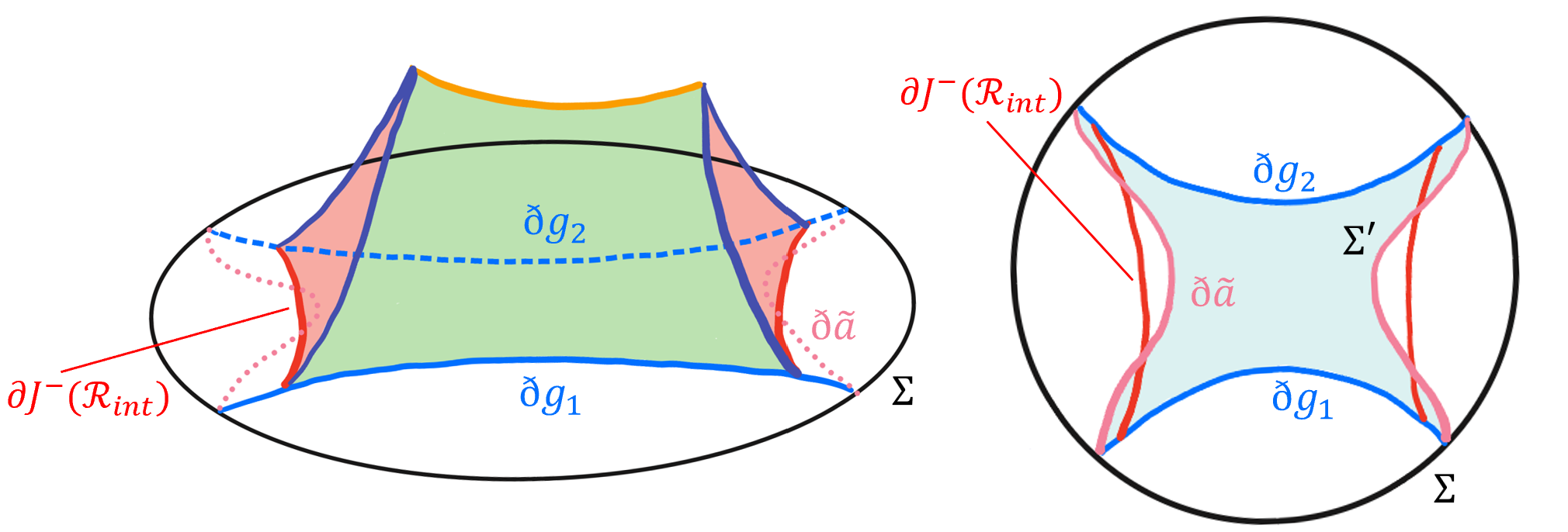}
    \caption{\textit{Left:}  Lift (green) and slope (red) used to find contradiction surface which is formed by the boundary of the past of the part of the ridge between the $b_{i}$, $\p J^{-}(\mathcal{R}_{int})$.     \textit{Right:} View of Cauchy slice $\Sigma$, including the edge of the fundamental complement $\eth a$. As depicted, we do not have enough information to determine if $\p J^{-}(\mathcal{R}_{int})$ is contained in $\Sigma'$ (shaded in light blue)
    } 
    \label{fig:diamondsslice}
\end{figure}

Rather than using $\p J^-(r_i)$, we have more control over the topology of the slope surface for this bulk version if we start directly from the segment of the ridge $\cal R$ between the $b_i$, which we'll denote ${\cal R}_{int}$. The boundary of the past of this surface $\p J^-({\cal R}_{int})$ is composed of part of the past lightcones of the $b_{i}$ and continuations of the generators of  $\p J^-(r_i)$ between the $b_{i}$. Because $\eth v_{i} =  \partial J^{+}(c_{i}) \cap \partial J^{-}(b_{i})$, $\p J^-({\cal R}_{int})$ will intersect $\eth v_{i}$. If we focus back along the generators of $\p J^-({\cal R}_{int})$ spanning the segment of the ridge $\cal R$ between the $b_i$ towards a Cauchy slice containing the $\eth g_i$ we will again be able to construct a contradiction surface as in the left hand side of Figure~\ref{fig:diamondsslice}.

However, $h \subset \tilde a'$ is not obviously true.
While we note that the the future boundary of the points based scattering region lives on the curve ${\cal R}$, it is not clear that this scattering region will be spacelike from the fundamental complement of $a=v_1\Cup v_2$. The intersection of $a'$ with the conformal boundary is shown in Figure \ref{fig:diamondsslice}. As before, $a' \cap \partial M = \hatit{X}_{1} \cup \hatit{X}_{2}$ and we let $k_{i}$ be the past boundary of $\hatit{X}_{i}$. However, except in the case of pure AdS,\footnote{Note that in global AdS there are no caustics. The fact that we know all of the generators from the $r_i$ converge to the same point $k_i$ at the conformal boundary tells us that the interior of the scattering region will be spacelike to the conformal shadow in this case. }  $k_{1}$ now lies to the past of the antipodal point of $r_{2}$. Additionally, the past directed boundary null geodesics hit the Cauchy slice in the interior of $\tilde{a}$. Even when we restrict to the $\p J^-({\cal R}_{int})$, which is all we would actually need to show is spacelike from the fundamental complement, we do not seem to have the necessary control over where the past lightcones from the $b_i$ versus from the rest of the ridge segment hit the boundary, and hence cannot shown that they hit $\Sigma$ outside of the fundamental complement.

Hence, by the arguments above, we can now claim that
\be\label{sptsconn2}
s_{pts}\neq\emptyset~\Rightarrow~ e_{min,0}(v_1\Cup v_2)~{\rm connected}
\ee
for $v_i$ given in~\eqref{vid}, assuming the $b_i$ are chosen so that the diamonds are non-empty and mutually spacelike. Indeed we can relax the latter condition by noting that if the $v_i$ are not spacelike the wedge union $v_1\Cup v_2$ will already be connected since any timelike curve between points in a wedge must remain inside the wedge so if the $v_i$ are not spacelike separated we can find a timelike curve between points in each. Then because $v_i\subset v_1\cup v_2$,
\be
e_{min,0}(v_1)\cup e_{min,0}(v_2)\subset e_{min,0}(v_1 \Cup v_2)
\ee
but since the right hand side is a wedge and $e_{min,0}(v_1)\Cup e_{min,0}(v_2)$ is the smallest wedge containing the left hand side we have
\be
e_{min,0}(v_1)\Cup e_{min,0}(v_2)\subset e_{min,0}(v_1 \Cup v_2).
\ee
The left hand side is connected by the same argument as above for $v_i$ not spacelike. Then $e_{min,0}(v_1\Cup v_2)$ will also be connected.

Note again that unless we have control over the other $e_{min,0}$ candidates our proof does not give a specific lower bound on the mutual information. However, we have thus have the desired result~\eqref{sptsconn2}. While we needed to resort to the weakened definition from~\cite{Bousso:2023sya} for this bulk version, many desirable properties of the min-entanglement wedges still hold. We will comment more on this in the discussion section~\ref{sec:disc}, as well as revisit the fate of $h \subset \tilde a'$ in the $b_i\rightarrow \p \tilde M$ limit in flat space in section~\ref{sec:flat}.

\subsection{A Connected Wedge Theorem for Screen Regions}

\begin{figure}
    \centering
    \includegraphics[width=0.5\linewidth]{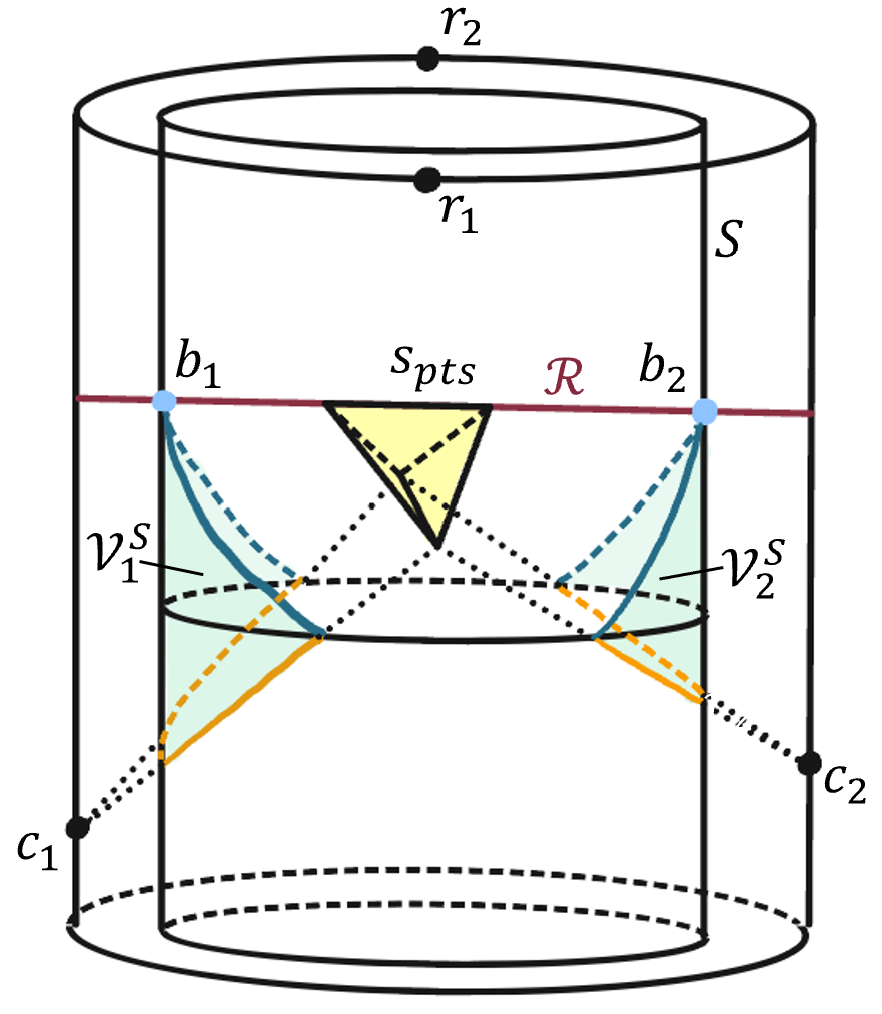}
    \caption{The cross sections $\mathcal{V}_{i}^{S}$ on a timelike screen $S$ and the scattering region $s_{pts}$.} 
    \label{fig:screen}
\end{figure}

We will now turn to a further generalization of the example from last section, where we will set up a bulk analog of the connected wedge theorem for regions on a timelike screen that avoids the $2\rightarrow 2$ scattering region.\footnote{We thank Chris Waddell for encouraging us to look for a statement about non-boundary anchored bulk regions and Alex May for prompting the generalization of section~\ref{sec:pairs} to the screen version considered here.}   Let $S$ denote our timelike screen, and let us define the screen regions
\be
{\cal V}^S_i= J^+(c_i)\cap J^-(r_1)\cap J^-(r_2)\cap S
\ee
as analogs of the boundary $\hatit{V}_i$ (see Figure \ref{fig:screen}). Now let $v_i$ denote the bulk casual wedge 
\be
v_i=({\cal V}^S_i)''
\ee
of the ${\cal V}^S_i$ in the ambient bulk. For simplicity, we will assume that our screen is such that each of the ${\cal V}^S_i$ consist of a single connected component, as one would expect for example a screen at fixed radial cutoff outside the scattering region. The claim is that~\eqref{sptsconn2} holds for this choice of $v_i$. First, note that the future boundary of $\mathcal{V}_{i}^{S}$ is a point $b_{i} \in J^-(r_1)\cap \p J^-(r_2)  \cap J^{+}(c_{i})$ like in the previous example. Note that 
\be
{\rm int}[J^+({\cal V}^S_i)\cap J^-({\cal V}^S_i)]\subset  v_{i} .
\ee 
Hence, the edge $\eth v_{i}$ is given by:
\begin{equation}
    \eth v_{i} = \partial J^{+}(\mathcal{V}_{i}^{S}) \cap \partial J^{-} (b_{i}) .
\end{equation}

In particular the part of $\p J^+( v_i)$ inside the screen will include some segment of the congruence from $c_i$. It is this segment that we will have control over the sign of the expansion for the focusing argument we used before. 

Now we need to show that this is the only segment relevant for our lift/slope surface. Indeed we want something slightly stronger if we want to show~\eqref{sptsconn2} and not just for some smaller scattering region $J^+(v_1)\cap J^+(v_2)\cap J^-(r_1)\cap J^-(r_2)\subset s_{pts}$. The first observation is that
\be
s_{pts}\subset J^+(c_i)\cap J^-(r_1)\cap J^-(r_2) .
\ee
Now we also know that ${\cal V}^S_i$ gives a cross section of this $1\rightarrow 2$ region.  If we could show that
\be
  J^+(c_i)\cap J^-(r_1)\cap J^-(r_2)\cap {\rm int }[S]\subset J^+({\cal V}^S_i)
\ee
we'd be done since by assumption our screen contains the scattering region. 
This will be true so long as the congruences from $c_i$ and $r_i$ intersect along a continuous spacelike curve that doesn't `double back.' 
If it did there would be multiple generators of $\p J^-(r_j)$ hitting the same generator of $\p J^+(c_i)$. But that would mean two points on different generators of $\p J^-(r_j)$ are lightlike separated. Then there would be a way to deform the concatenation to a timelike curve contradiction the past point being on the light cone.\footnote{\label{Jmri}
Note that we only needed properties of $\p J^-(r_i)$ for this argument and would have reached the same conclusion about $\p J^-(r_i)\cap \p J^+({\cal E})$ for ${\cal E}$ some extremal surface. This assumption is also built into our discussions about the topology of $s_{pts}$ in figure~\ref{fig: spts and sreg} and footnote~\ref{ft:sptstop}.
}

Now we see that $\eth v_i$ forms a cross section of the segment of $\p J^+(c_i)$ that contributes to the $1\rightarrow 2$ bulk scattering region and as we follow these generators towards the scattering region they either hit segment of $\p J^+(c_i)\cap \p J^+(r_j)$ or the ridge. By the comment in footnote~\ref{Jmri}, we have all the ingredients we need to follow the proof from the previous section starting from the ridge segment ${\cal R}$ between the points $b_i$ where it hits the screen in figure~\ref{fig:screen}, which we will denote ${\cal R}_{int}$ as for the previous configuration. As in the previous causal diamond example, we have limited control over $\tilde a$, and so can prove the weaker version~\eqref{sptsconn2}. 

\vspace{1em}\noindent 
Before turning to the flat space version, let us make one further observation. Consider a Cauchy slice $\Sigma$ containing $\eth v_i$ and 
$\eth e_{min,0}(v_1\Cup v_2)$ satisfying the third condition in Definition~\ref{def:min0}. 
If the portion of $\p J^-({\cal R}_{int})\cap \Sigma$ between where the congruence hits the $\eth v_i$
was outside of the wedge, we know that $\p J^-({\cal R}_{int})$ will intersect $\eth e_{min,0}(v_1\Cup v_2)$. 
Now recall from Theorem~\ref{minedge} (which also holds for $e_{min,0}$~\cite{Bousso:2023sya}) that the edge of the wedge will have a vanishing expansion along the outward future normal if the segment is extremal or on $H^-(a)$. If instead it is on $H^+(a)$, then it will also be non-positive due to the fact that, by assumption we are inside $J^-(r_1)\cap J^-(r_2)$ so these segments of $H^-(a)$ are generated by $\p J^+(c_i)$. Then if I focus up from this segment of the wedge and down along $\p J^-({\cal R}_{int})$ we see that the segment is smaller, and so we again get a contradiction to condition 3. From this 
we can conclude that $\p J^-({\cal R}_{int})\cap \Sigma$ is inside the entanglement wedge. Moreover the  $\p J^-({\cal R}_{int})$ will contain the scattering region (see footnote~\ref{ft:sptstop} and note ${\cal R}_{int}$ contains the future outgoing ridge of the scattering region)  so
\be
s_{pts}\subset e_{min,0}(v_1\Cup v_2).
\ee
Furthermore by symmetry this implies 
\be
s_{pts}\subset e_{min,0}(v_1\Cup v_2)\cap e_{min,0}(w_1\Cup w_2)=:s_{ent,0}
\ee
where the $w_i$ are outgoing analogs. We thus have the same nesting structure for the points based and entanglement scattering regions in the bulk story as we had in section~\ref{sec:scatreg}. Note that this will hold so long as the segment of $\p J^-({\cal R}_{int})$ from the future ridge of the scattering region is necessarily inside $e_{min,0}(v_1\cup v_2)$. This followed for our screen region since by assumption the screen is spacelike to the scattering region. It would also hold for the example in section~\ref{sec:pairs} if we restrict to pairs of diamonds that are mutually spacelike.

\section{Scattering and Entanglement in Flat Space}\label{sec:flat}

The goal of the previous section was to construct decision regions that were amenable to the flat space limit. Let us now examine the flat limit to see why this was necessary and what statements carry over. While we could jump directly to asymptotically flat spacetimes (AFS), it will be more illustrative to look at the bulk point limit to see how the AdS scale resolves things. Let's start with global AdS$_3$ for simplicity, though the general aAdS$_3$ to AFS$_3$ limit is straightforward and can be found in e.g.~\cite{Barnich:2012aw}. Starting with the metric
\be\label{ads}
ds^2=-\left(1+\frac{r^2}{\ell^2}\right)\ell^2 d\tau^2+\frac{1}{1+\frac{r^2}{\ell^2}}dr^2+r^2 d\phi^2 \ee
the coordinate transformation \cite{Compere:2020lrt}
\be
\tau=\frac{u}{\ell}+\arctan\frac{r}{\ell}
\ee
takes us to
\be\label{metlim}
ds^2=-\left(1+\frac{r^2}{\ell^2}\right)du^2-2dudr+r^2 d\phi^2.
\ee
The $\ell\rightarrow\infty$ limit clearly takes us to Minkowski space. A similar coordinate transformation can be used to map to advanced coordinates $(v,r,\phi)$. Now $(\tau,\phi)$ in~\eqref{ads} are coordinates on the boundary $\mathbb{R}\times S^1$.
If we take the input points $c_i$ to lie near the band $\tau =- \frac{\pi}{2}+\frac{v}{\ell} $ and output points $r_i$ near $\tau=\frac{\pi}{2} +\frac{u}{\ell}$ before sending the AdS scale to infinity, we will land on scattering process from $\scri^-$ to $\scri^+$ in flat space from this bulk point configuration. 

Now let's examine what happens to the decision regions in this limit. Staring at~\eqref{metlim} we see that at large $r$ the metric is
\be
ds^2\simeq r^2(-\frac{1}{\ell^2}du^2 + d\phi^2)
\ee
so that the flat limit is effecting a Carrollian limit of the boundary where the speed of light $c=\frac{1}{\ell}\rightarrow 0$ goes to zero~\cite{Levy-Leblond:1965dsc,Duval:2014uoa}. This effectively collapses the lightcone to a single generator. If we construct the boundary input regions
\be
\hatit{V}_i =\hat J^+(c_i)\cap \hat J^-(r_1)\cap \hat J^-(r_2)
\ee
for $c_i(v_{c_i},\phi_{c_i})$ and $r_j(u_{r_j},\phi_{r_j})$, 
we see that the left and right corners of these diamonds are given by 
\be\label{v1}
\badat{3}
(\tau,\phi)=&\{(\frac{u_{r_1}+v_{c_1}+\ell(\phi_{c_1}-\phi_{r_1})}{2\ell},\frac{-u_{r_1}+v_{c_1}+\ell (-\pi+\phi_{c_1}+\phi_{r_1})}{2\ell}),~ \\
&(\frac{u_{r_2}+v_{c_1}+\ell (-\phi_{c_1}+\phi_{r_2})}{2\ell},\frac{u_{r_2}-v_{c_1}+\ell(\pi+\phi_{c_1}+\phi_{r_2})}{2\ell})\}
\eadat
\ee
for $\hatit{V}_1$ and
\be\label{v2}
\badat{3}
(\tau,\phi)=&\{(\frac{u_{r_2}+v_{c_2}+\ell(\phi_{c_2}-\phi_{r_2})}{2\ell},\frac{-u_{r_2}+v_{c_2}+\ell (-\pi+\phi_{c_2}+\phi_{r_2})}{2\ell}),~ \\
&(\frac{u_{r_1}+v_{c_2}+\ell (-\phi_{c_2}+\phi_{r_1})}{2\ell},\frac{u_{r_1}-v_{c_2}+\ell(\pi+\phi_{c_2}+\phi_{r_1})}{2\ell})\}
\eadat
\ee
for $\hatit{V}_2$. Here we note $\phi\sim\phi+2n\pi$. Then in the $\ell\rightarrow\infty$ limit these degenerate to the semi-infinite segments 
\be\label{flatVi}
 \hatit{V}_{1} = \{ (v, \phi) : v \geq v_{c_{1}}, \phi = \phi_{c_{1}}\},~~ \hatit{V}_{2} = \{ (v, \phi) : v \geq v_{c_{2}}, \phi = \phi_{c_{2}}\}
\ee
on $\scri^-$ while the left and right end points in~\eqref{v1}-\eqref{v2} become
\be\label{flatv12}
\badat{3}
(\tau,\phi)=&\{(\frac{1}{2}(\phi_{c_1}-\phi_{r_1}),\frac{1}{2}(-\pi+\phi_{c_1}+\phi_{r_1})),~\\
&(\frac{1}{2}(-\phi_{c_1}+\phi_{r_2}),\frac{1}{2}(\pi+\phi_{c_1}+\phi_{r_2}))\},
\eadat
\ee
and
\be\label{flatv22}
\badat{3}
(\tau,\phi)=&\{(\frac{1}{2}(\phi_{c_2}-\phi_{r_2}),\frac{1}{2}(-\pi+\phi_{c_2}+\phi_{r_2})),~ \\
&(\frac{1}{2}(-\phi_{c_2}+\phi_{r_1}),\frac{1}{2}(\pi+\phi_{c_2}+\phi_{r_1}))\} ,
\eadat
\ee
respectively, which we can think of as being a blow up of $i^0$.
The description of the $\hatit{W}_i$ at $\scri^+$ is analogous. 

\begin{figure}
    \centering
    \includegraphics[width=0.5\linewidth]{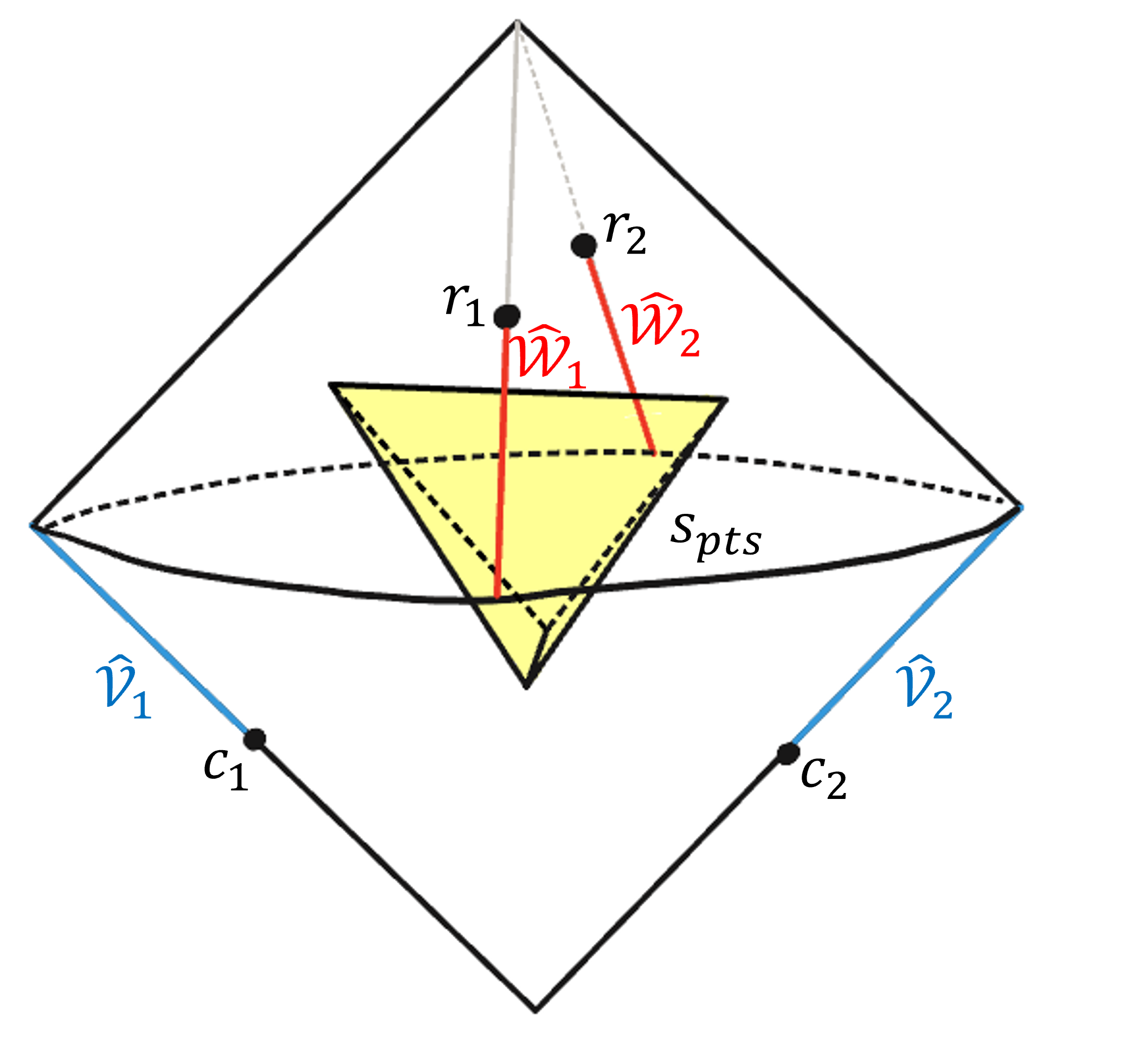}
    \caption{Scattering configuration in an asymptotically flat spacetime. The decision regions become degenerate on the boundary, while the bulk scattering region $s_{pts}$ remains sensitive to the positions of the input and output points.} 
\label{fig:flatscattering}
\end{figure}

In the BP proposal, one would typically only consider the boundary at $\scri^\pm$ for asymptotically flat spacetimes, with $i^{0}$ being a single point in the compactification. In this case we lose any information about the $r_i$ in the expressions for $\hatit{V}_i$ in~\eqref{flatVi}.  From~\eqref{flatv12}-\eqref{flatv22} we see that even if we tried to restore the angular data at $i^0$, we only regain the angular information about the outgoing points. Meanwhile we know that the scattering region in the bulk depends sensitively on both the $c_i$ and $r_i$ positions. Concretely, in the case of $\mathbb{R}^{1,2}$ the scattering region is just carved out by null hyperplanes which we can see as follows. Writing 
\be
X^\mu =rq^\mu +u t^\mu
\ee
 where $t^\mu=\delta^\mu_0$ is a reference timelike vector and $q$ a  null vector we see that in the limit we send a reference point $X_0$ to $\scri^+$ by taking $r_0\rightarrow\infty$ we get
\be\badat{3}\label{eq:scriplus}
(X-X_{0})^2 &\simeq  -2r_0  q_0\cdot ( X-u_0t)
\eadat\ee
where we've suppressed corrections that are higher order in $r_0$. We thus see that $q_0$ sets the direction of the null hyperplane while $u_0$ fixes the impact parameter. Applying this to our $2\rightarrow 2$ scattering problem, we see that the existence of bulk scattering is highly sensitive to both $\phi_{r_i}$ and $u_{r_i}$, while the boundary input regions~\eqref{flatVi} lose some of this data.

We thus see why we were motivated to look for generalizations of the CWT like in section~\ref{sec:bulkcwt}. While the scattering region naturally survives the bulk point limit, the boundary decision regions (and RT surfaces for that matter~\cite{Ghosh:2023kuy,Jiang:2017ecm,jmp,cmp}) degenerate. These kinematical issues are resolved by introducing a screen in flat space, but then we want to make sure we have a version of the CWT that still holds.\footnote{Note from here on we are considering a screen directly in an asymptotically flat spacetime. Alternatively, we could also imagine considering our $c_i$ and $r_i$ to be in a  bulk point configuration in AdS, and then introducing a screen or (choice of $b_i$) that goes very deep into the bulk, surrounding a sub-AdS scale region~\cite{Lewkowycz:2019xse}, before proceeding to take a flat limit.
\label{ft:AdSbpscreen}} Fortunately, the generalized entanglement wedge construction of Bousso and Penington 
does not depend on the value of the cosmological constant of the ambient bulk. Because our regions still involve $c_i$ and $r_i$ at the conformal boundary, we still have the conditions on the expansions for those lightsheets -- and thus the version \eqref{sptsconn2} of our bulk CWT with $e_{min,0}$ -- should carry over to asymptotically flat space.\footnote{Note that one concern with the definition of $e_{min,0}$ in~\cite{Bousso:2023sya} is whether or not one is guaranteed existence given the restriction $\tilde\eth g=\tilde \eth a$.  
Or that $e_{min,0}$ could be whole spacetime already with one region, which while in principle could still be consistent with a connected wedge theorem, would make the implication a bit vacuous. This motivated the modified definition used in~\cite{Bousso:2025fgg} when the authors were interested in generalizing beyond asymptotically AdS spacetimes. Here we note that for the particular regions of interest the intersection of the decision regions with $\scri^-$ form a single ray so that  $\tilde \eth a$ degenerate to the same point with opposite orientation. This would seem to help evade the aforementioned issues. We thank Sami Kaya for discussions on this point. }

Let's close by considering what happens when our screen approaches the boundary or, equivalently, $b_i\rightarrow \tilde \p M$ in the set up from section~\ref{sec:pairs}. As discussed above, going all the way to the boundary puts us in the situation where the regions $\hatit{V}_i$ lose some information about the scattering configuration, but in the limit that the cutoff is large but finite we can ask if something nice happens for property 1. We will use the diamond configuration from section~\ref{sec:pairs} for concreteness since it's simpler. For aAdS spacetimes we saw that the large angular momentum geodesics were insensitive to the bulk geometry since they only saw the asymptotic region. This feature helped us prove the stronger version~\eqref{sptsconn} of the bulk hologram CWT at the beginning of section~\ref{sec:bulkcwt}, since we had control over the lightsheets from the $r_i$ and $\cw{\hatit{X}_{i+1}}$ regions. 
So long as this remains true for the large angular momentum generators in AFS spacetimes we could hope that in the limit where we send $b_i\rightarrow \tilde \p M$ the past light cones from $b_i$  would approach hitting $\scri^-$ at the same point.\footnote{While it is tempting to co-opt the AdS result following the logic in footnote~\ref{ft:AdSbpscreen}, one should be careful about the order of limits between sending the impact parameter versus the AdS radius large.} As a warm up, let's see what happens in Minkowski space.

In Minkowski space, the past lightcone of the $r_i$ only hits $\scri^-$ at the antipodal point $v=u_{r_i}$, $\phi=\phi_{r_i}+\pi$. Let us see this in equation. First, we can adapt~\eqref{eq:scriplus} to $\scri^-$ via $X_0^\mu=-r_0q_0^\mu+v_0 t^\mu $. Then the past lightcone of a bulk point hits $\scri^-$ when 
\be
0=q_0\cdot(X-v_0t).
\ee
Note that if $X$ were just the origin this sets $v_0=0$, as expected for the lightcone of the origin.
To send $X$ towards spatial infinity it is useful to use the hyperbolic coordinates 
\be
X=\rho(\sinh\eta,\cosh\eta\cos\phi,\cosh\eta\sin\phi).
\ee
Then we have 
\be\label{solvev0}
0=v_0+\rho(\cos(\phi-\phi_0)\cosh\eta-\sinh\eta).
\ee
If we just naively sent $\rho\rightarrow\infty$ holding the other variables constant, we'd get $\phi_0$ fixed to a pair of values determined by the direction  $(\eta,\phi)$ on the blowup of spatial infinity, and $v_0=0$. However, once we also ask for this point to be on the past lightcone of the $r_i$ we have additional constraints
\be
0=q_{r_i}\cdot (X-u_{r_i}t)
\ee
which we can use to solve for $\eta$ and $\phi$, which give them $1/\rho$ corrections that will affect the value of $v_0$ at leading order when we solve~\eqref{solvev0} in a $1/\rho$ expansion.  Namely, we recover the points $v_0=u_{r_i}$, $\phi=\phi_{r_i}+\pi$, as expected. We thus have two special generators on $\scri^-$ where the lightcones from the $b_i\rightarrow\tilde \p M$ hit the conformal boundary. Meanwhile all generators of $\scri^-$ closer to $\phi_{b_1}$ will be timelike from $b_1$, and further from $\phi_{b_1}$ will be spacelike from it. The same goes for $b_2$. The intersection of $(v_{1} \Cup v_{2})'$ with the conformal boundary is $\hatit{X} = \hatit{X}_{1} \cup \hatit{X}_{2}$ with: 
\be\label{flatXi}
 \hatit{X}_{1} = \{ (v, \phi) : v \geq u_{r_{2}}, \phi = \phi_{r_{2}}+\pi\},~~ \hatit{X}_{2} = \{ (v, \phi) : v \geq u_{r_{1}}, \phi = \phi_{r_{1}}+\pi\}.
\ee
As in the global AdS case, the scattering region is thus clearly spacelike from the fundamental complement in Minkowski space.

We also see that if it were true that, as in aAdS geometries, the large angular momentum null geodesics in AFS spacetimes converge to the antipodal points on the conformal boundary we expect the conformal shadow of the union of the causal diamonds $v_i$ regions to just include these segments. Then by the same time delay argument for the bulk geodesics we'd be guaranteed that the scattering region is spacelike from $\tilde a$.\footnote{Note that~\cite{Bousso:2025fgg} actually contains two definitions of the fundamental complement. The one above requires the timelike geodesics to have infinite extent to both the future and past, which would make the causal wedge of the $\hatit{X}_i$ trivial, while the one in their appendix requires it only need by semi-infinite. } Note that one should really be careful about statements that hold in the strict $b_i\rightarrow \tilde\p M$ limit where the input regions are also degenerate, however it is instructive to see what features for the AFS geodesics we would need to reproduce the results for AdS.

\section{Discussion}\label{sec:disc} 
In this paper we have used the framework of generalized entanglement wedges~\cite{Bousso:2022hlz,Bousso:2023sya} to understand how to export the results of the connected wedge theorem~\cite{May:2019odp} to asymptotically flat spacetimes. In the process, we established new upper and lower bounds for the original AdS result in terms of the entropies of bulk scattering regions. We also identified bulk generalizations of the decision regions in~\cite{May:2019odp} which obey a connected wedge theorem for the min entanglement wedges defined in~\cite{Bousso:2023sya}.

Some natural follow up directions include looking at how to extend this to the $n\rightarrow n$ processes~\cite{May:2022clu},\footnote{See also~\cite{Zhao:2025rnm} for an exploration of $n\rightarrow n$ in the context of a GCWT with converse~\cite{Leutheusser:2024yvf,Lima:2025dtj,Zhao:2025sso}.} as well as seeing if this machinery helps us generalize the connected wedge theorem to higher dimensions. It would also be fruitful to understand what happens when we add matter, so that the bulk entropies are no longer just area terms, as well as to make contact with other investigations of entanglement entropy in asymptotically flat spacetimes~\cite{Jiang:2017ecm,Li:2010dr,cmp,jmp}. While these sorts of continuations chart several possible next steps forward, 
let us close by commenting on some deeper conceptual questions that emerge from this framework.

First, our investigations in section~\ref{sec:bulkcwt} seem to indicate there is more to explore with regards to which of the various refinements of the proposals for bulk entanglement wedges (e.g.~\cite{Bousso:2023sya} vs~\cite{Bousso:2025fgg}) are physically relevant. For example, while the fundamental complement in~\cite{Bousso:2025fgg} gives an analog of complementarity which seems rather desirable, many of the other features we would want for an entanglement entropy are still satisfied by the original definition~\cite{Bousso:2023sya}. This begs the question of whether different holograms apply in different contexts, or whether there are further alternatives to explore. If our aim is solely to recover flat space results when there are large cutoffs, we might be less concerned about using the relaxed definition in intermediate steps in section~\ref{sec:bulkcwt}. Nevertheless, our findings seem to indicate that scattering protocols might help differentiate these variants. 

Second, it is worth pointing out that the mechanism by which we introduce the screen is different from what one would do in the $T\bar T$ deformation story~\cite{Li:2010dr,Lewkowycz:2019xse} or in the context of holographic tasks in de Sitter in~\cite{Franken:2024wmh}. In particular, the notion of using the bulk entanglement wedge in the ambient space rather than anchoring the minimal surfaces to the boundary is more agnostic to how we would otherwise try to tie the screen to the boundary, and the particularities of the embedding of the screen. We hope to see how this can be useful in other contexts.

Finally, let us return to an interpretation of the bounds in section~\ref{sec:scatreg}. As discussed in section~\ref{sec:prelim}, 
the original inspiration for the connected wedge theorem~\cite{May:2019yxi} came from  quantum tagging results that showed how a sufficient amount of entanglement allows you to spoof scattering through an excluded region \cite{Kent2006TaggingSystemsGranted,Kent_2011, Buhrman_2014}. On the QI side the lower bound on the mutual information is linear in the number of bits you are trying to send. Now that we have a scattering region entropy interpretation of the lower bound, we can try to see how it fits into the picture. Indeed, it seems natural that the amount of information contained in the scattering region is related to the amount of information we can send in the protocol.\footnote{The authors of~\cite{May:2019odp} also considered interpreting the ridge area in a similar manner via the Bousso bound. We are grateful to Alex May for discussions on this point.}

It would be worth making these statements more precise. In particular, now that non-local protocol dual has also been moved into the bulk, combining the discussion of screens, quantum tagging, and adding matter terms might help clarify exactly how these seemingly purely geometric statements are concretely implied by the quantum tasks. As these discussions have hopefully illustrated, there are many facets of the generalized entanglement wedge proposal worth exploring and ample opportunities for future work.

\section*{Acknowledgments}

We would like to thank Raphael Bousso, Jackie Caminiti, Sami Kaya, Caroline Lima, Alex May, Takato Mori, Rob Myers, Elisa Tabor, and Chris Waddell for helpful discussions. This work was supported by the Celestial Holography Initiative at the Perimeter Institute for Theoretical Physics and the Simons Collaboration on Celestial Holography.  Research at the Perimeter Institute is supported by the Government of Canada through the Department of Innovation, Science and Industry Canada, and by the Province of Ontario through the Ministry of Colleges and Universities.

\appendix

\section{Proving the Ridge Upper Bound}\label{app:ridge}

In this appendix we will write out the proof of~\eqref{ridgeupper} 
\begin{thm}
\label{thm: gwct}
 $I(\hatit{V}_{1}; \hatit{V}_{2}) \leq \sum_{i=1}^2 \frac{1}{4 G_{N}}\area{\mathcal{R}_{ent,i}}$ .
\end{thm}
\noindent 
Again we will be assuming $s_{pts}$ is non-empty so that the wedges are guaranteed to be connected. Recall that the composite ridge was defined in~\eqref{eq: Rei} as 
\begin{equation}
    \mathcal{R}_{ent,i} =  {\cal R}_+\cup {\cal R}_-\cup  \mathcal{S}_{i}
\end{equation}
where $r_\pm,~{\cal S}_i$ are given in~\eqref{eq: rp}-\eqref{eq:si}. As discussed in section~\ref{sec:upper}, while the proof of~\eqref{sentupper} involved focusing from the edge of the wedge $\eth s_{ent}$ towards the ridge, the proof of Theorem~\ref{thm: gwct} here will involve focusing from the ridge towards $\Sigma$. 

\begin{figure}
    \centering
    \includegraphics[width=0.9\linewidth]{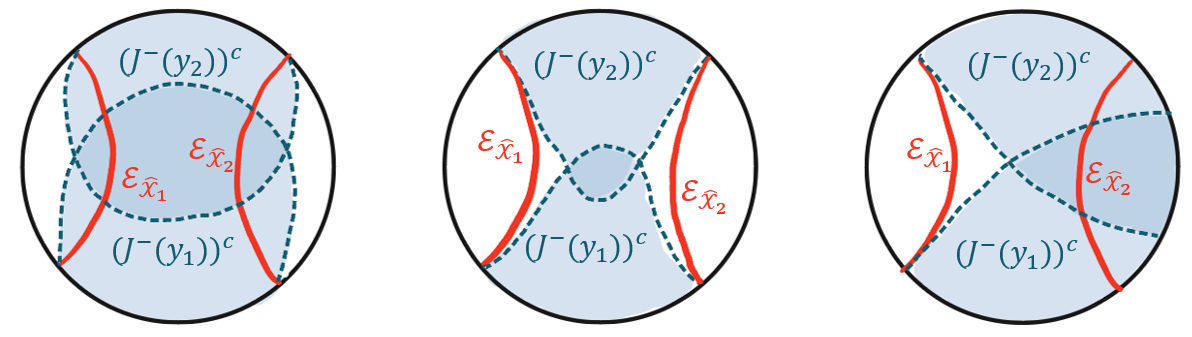}
    \caption{The Cauchy slice $\Sigma$ illustrated for three possible cases. The regions $(J^{-}(y_{1}))^{c}$ and $(J^{-}(y_{2}))^{c}$ are shaded in blue and their boundaries are indicated with a dashed blue line. Wherever these two regions intersect, is part of the scattering region $s_{ent}$. 
    }
    \label{fig:pf cases}
\end{figure}

We can see from the definitions that $r_\pm$ hit $\Sigma$ precisely at the edges of the ${\cal S}_i$. Combined with a result from~\cite{Lima:2025dtj} that $s_{ent}$ `cuts across' $\ew{\V_1\cup \V_2}$, we have that the ridges $ \mathcal{R}_{ent,i}$ will form a continuous curve that crosses between the part of $H^\pm(\ew{\V_1\cup \V_2})$ that overlaps with $H^\pm(\ew{X_1}')$ and that overlaps with $H^\pm(\ew{X_2}')$. We can thus form a lift and slope surface analogous to~\eqref{N1} and~\eqref{N2}, but now we will again split things into parts above and below $\Sigma$:  
\begin{equation}\label{N1p}
    \mathcal{N}_{1}^+ = \partial \left[ J^{-}(y_{1}) \cup  J^{-}(y_{2})\right] \cap (J^{+}(x_{1}))^{c} \cap (J^{+}(x_{1}))^{c} \cap J^{+}(\Sigma)
\end{equation}
\begin{equation}\label{N1m}
    \mathcal{N}_{1}^- =\partial \left[ J^{+}(y_{1}) \cup J^{+}(y_{2})\right] \cap (J^{-}(x_{1}))^{c} \cap (J^{-}(x_{1}))^{c} \cap J^{-}(\Sigma)
\end{equation}
and
\begin{equation}\label{N2p}
    \mathcal{N}_{2}^+ =  \partial \left[J^{+}(x_{1}) \cup  J^{+}(x_{2})\right] \cap J^-(y_1)\cap J^-(y_2)
\end{equation}
\begin{equation}\label{N2m}
     \mathcal{N}_{2}^- =  \partial \left[J^{-}(x_{1}) \cup  J^{-}(x_{2})\right] \cap J^+(y_1)\cap J^+(y_2).
\end{equation}
While these definitions look slightly different than the notation in~\cite{Lima:2025dtj}, one can verify that they are equivalent for the case considered there. 
Then ${\cal N}^\pm_1$ meets $\Sigma$ at $\mathcal{D}^\pm =\mathcal{D}^\pm_{1} \cup \mathcal{D}^\pm_{2}$ where:
\begin{equation}
    \mathcal{D}_{i}^+ = \partial J^{-}(y_{i})  \cap (J^{+}(x_{1}))^{c} \cap (J^{+}(x_{1}))^{c} \cap \Sigma
\end{equation}
\begin{equation}
    \mathcal{D}_{i}^- = \partial J^{+}(y_{i})  \cap (J^{-}(x_{1}))^{c} \cap (J^{-}(x_{1}))^{c} \cap \Sigma .
\end{equation}
Meanwhile ${\cal N}^\pm_1$ and ${\cal N}^\pm_2$ meet at seams ${\cal A}^\pm$. These seems hit $\Sigma$ at the end points of the ${\cal D}_i$. These will either be at some finite segment of the ${\cal E}_{\hat \X_j}$ or at their end points. Some examples for how the ${\cal D}_i$ intersect the ${\cal E}_{\hat\X_i}$ are illustrated in figure~\ref{fig:pf cases}. 
Let 
\begin{equation}
    \mathcal{C}_{i} = \E{\hatit{X}_{i}} \cap s_{ent}
\end{equation}
and let $\mathcal{A}_{ij}$ denote the pieces of the extremal surfaces which are not contained in $s_{ent}$:
\begin{equation}
    \mathcal{A}_{ij} = (\E{\hatit{X}_{i}} \cap (J^{-}(y_{j}))^{c}) - \mathcal{C}_{i}.
\end{equation}
Then the curve
\be
{\cal F}_i={\cal \D}_i \cup \mathcal{A}_{1i}\cup \mathcal{A}_{2i}
\ee
is homologous to $\V_i$ and has larger area
\be\label{F}
\sum_{j=1}^2 \area{{\cal A}_{ij}}+\area{{\cal D}_i} =\area{{\cal F}_i}\ge \area{\V_i}
\ee
while focusing tells us 
\be\label{dcr}
\sum_{i=1}^2 [\area{{\cal \D}_i}-\area{{\cal C}_i}]\le \sum_{i=1}^2 \area{\mathcal{R}_{ent,i}}.
\ee
Now we also have
\be\label{X}
\area{\hat {\cal X}_i}= \sum_{j=1}^2 \area{{\cal A}_{ij}}+\area{{\cal C}_i}
\ee
so that, combining~\eqref{F} and~\eqref{X} we get
\be\badat{3}
\area{\V_i}-\area{\hat {\cal X}_i} 
\le \area{{\cal D}_i}-\area{{\cal C}_i}.
\eadat\ee
Finally~\eqref{dcr} tells us
\be
I(\hatit{V}_{1}; \hatit{V}_{2}) \leq \sum_{i=1}^2 \frac{1}{4 G_{N}}\area{\mathcal{R}_{ent,i}}
\ee
proving Theorem~\ref{thm: gwct}.

\bibliographystyle{JHEP}
\bibliography{refs}

\end{document}